\documentclass[proceedings, preprint]{rmaa}
\usepackage{paralist}
\usepackage{psfrag,color}
\usepackage{natbib}
\usepackage{color}
\usepackage[colorlinks=true,citecolor=blue, linkcolor=blue, urlcolor=blue]{hyperref}
\usepackage{cuted}
\usepackage{etoolbox}
\usepackage{xspace}
\usepackage{float}
\usepackage{placeins}
\usepackage{threeparttable}
\usepackage{tablefootnote}
\usepackage{longtable}
\AfterEndEnvironment{strip}{\leavevmode}

\newcommand{\thisgrb}{GRB~210210A\xspace}

\newcommand{\kw}{{\em Konus}-Wind\xspace}

\newcommand{\keV}{{\rm keV}\xspace}

\newcommand{\swift}{{\em Swift}\xspace}
\newcommand{\tninty}{{$T_{\rm 90}$}\xspace}
\newcommand{\tzero}{T$_{\rm 0}$\xspace}

\newcommand{\Ep}{$E_{\rm p}$\xspace}
\newcommand{\sw}[1]{\texttt{#1}}

\usepackage{xcolor}

\SetYear{2023}
\SetConfTitle{7th Workshop on Robotic Autonomous Observatories}
\title{An Intermediate Luminosity \thisgrb: The early onset of the external forward shock in the X-ray?} 

\author{Rahul Gupta,\altaffilmark{1,2,3}
A. K. Ror,\altaffilmark{1,4}
S. B. Pandey,\altaffilmark{1} 
J. Racusin,\altaffilmark{2}
M. Moss,\altaffilmark{2,3}
A. Aryan,\altaffilmark{1,5}
N. Klingler,\altaffilmark{2}
and A. J. Castro-Tirado\altaffilmark{6,7} }

\altaffiltext{1}{Aryabhatta Research Institute of Observational Sciences (ARIES), Manora Peak, Nainital-263002, India}
\altaffiltext{2}{Astrophysics Science Division, NASA Goddard Space Flight Center, Mail Code 661, Greenbelt, MD 20771, USA}
\altaffiltext{3}{NASA Postdoctoral Program Fellow}
\altaffiltext{4}{Department of Applied Physics/Physics, Mahatma Jyotiba Phule Rohilkhand University, Bareilly-243006, India}
\altaffiltext{5}{Graduate Institute of Astronomy, National Central University, 300 Jhongda Road, 32001 Jhongli, Taiwan}
\altaffiltext{6}{Instituto de Astrofisica de Andaluc\'{i}a (IAA-CSIC), Glorieta de la Astronom\'{i}a s/n, E-18008, Granada, Spain}
\altaffiltext{7}{Unidad Asociada Departamento de Ingenier\'{i}a de Sistemas y Autom\'{a}tica, E.T.S. de Ingenieros Industriales, Universidad de M\'{a}laga, E-29071, Spain}

\shortauthor{R. Gupta et~al.}
\shorttitle{\thisgrb: An intermediate luminosity burst}

\abstract{We have analyzed the prompt and afterglow characteristics of the intermediate luminosity burst ``GRB 210210A". Our prompt emission analysis indicates that GRB 210210A is among the softest long GRBs detected by the \swift-BAT. The time-integrated prompt emission spectrum of GRB 210210A is aptly described by a power law function with an exponential cutoff. The spectral peak energy (E$_{p,z}$) in rest-frame and the E$_{\rm \gamma, iso}$ for this GRB marginally satisfy the 2$\sigma$ Amati correlation, a common feature observed in low/intermediate luminosity GRBs. Notably, an early bump is observed in the \swift-XRT light curve (a rare feature); the optical afterglow light curve, on the other hand, appears to follow a power law decay. However, due to the lack of sufficient early optical observations, we cannot completely rule out the possibility of an early bump in the optical light curve. For the bump observed in the early X-ray light curve, we calculated parameters such as peak time, rise time, decay time, and bulk Lorentz factor ($\Gamma_{0}$ $\sim$ 156), which perfectly satisfy the correlation between the parameters of the onset of the afterglow in GRBs. Both the optical and X-ray (including our observations) light curves exhibit a chromatic break in the late afterglow. Based on the prompt and afterglow parameters, we confirm that the intermediate luminosity GRB 210210A favors a collapsar scenario and is possibly powered by a magnetar.}

\resumen{Hemos analizado las carácter\'isticas de la emisi\'on de la posluminiscencia del estallido de rayos gamma de luminosidad intermedia GRB 210210A. Nuestro an\'alisis de la emisi\'on de la misma indica que GRB 210210A es uno de los GRB largos m\'as suaves detectados por {\it Swift}-BAT. El espectro de emisi\'on de dicha posluminiscencia integrado en el tiempo de la duraci\'on de GRB 210210A se describe adecuadamente mediante una funci\'on de ley de potencias con un corte exponencial. La energ\'ia del pico espectral (E$_{p,z}$) en el sistema de referencia de reposo y el valor de la energ\'ia (asumiendo isotrop\'ia) E$_{\rm \gamma, iso}$ para este GRB satisfacen marginalmente (2$\sigma$) la correlaci\'on de Amati, una característica com\'un observada en GRBs de luminosidad baja/intermedia. Cabe destacar que se observa un declive temprano en la curva de luz de {\it Swift}-XRT (una característica poco com\'un); mientras que la curva de luz de la posluminiscencia en el \'optico, por otro lado, parece seguir un decaimiento en forma de ley de potencias. Sin embargo, debido a la falta de suficientes observaciones \'opticas tempranas, no podemos descartar por completo la posibilidad de una protuberancia temprana en la curva de luz \'optica. Para la protuberancia observada en la curva de luz de rayos X temprana, calculamos par\'ametros como el tiempo de pico, el tiempo de ascenso, el tiempo de decaimiento y el factor de Lorentz ($\Gamma_{0}$ $\sim$ 156), que satisface perfectamente la correlaci\'on entre los par\'ametros del inicio de la posluminiscencia en los GRB. Tanto la curva de luz \'optica como la de rayos X (incluidas nuestras observaciones) exhiben una ruptura crom\'atica tard\'ia para dicha posluminiscencia. Con base en los parámetros de dicha protuberancia y postluminiscencia, confirmamos que la luminosidad intermedia del GRB 210210A favorece un escenario de colapso y que posiblemente sea debido a un magnetar subyacente.}

\addkeyword{gamma-ray burst: general}
\addkeyword{gamma-ray burst: individual (GRB 210210A)}
\addkeyword{methods: data analysis}
\addkeyword{stars: jets}
\addkeyword{techniques: photometric}

\begin{document}
\maketitle

\section{Introduction} 
\label{sec:intro}

\indent Gamma-ray bursts (GRBs) are among the most energetic and luminous events in the cosmos, typically observed as brief flashes of gamma-ray light (prompt emission), originating from distant galaxies. They are broadly classified into two categories based on their duration: long-duration GRBs, which endure for more than two seconds and are believed to result from the collapse of massive stars \citep{1998Natur.395..670G, 1999ApJ...524..262M, 2003Natur.423..847H, 2004RvMP...76.1143P}, while short-duration GRBs, generally lasting less than two seconds, are commonly attributed to the merger of compact binaries comprising either two neutron stars or a neutron star and a black hole \citep{2002ApJ...570..252P, 2017ApJ...848L..13A}. However, recently, a few outliers/hybrid events (long bursts from merger and short bursts from collapsar) have also been discovered \citep{2021NatAs...5..917A, 2022Natur.612..228T, 2022ApJ...931L..23L, 2024Natur.626..737L}.

\noindent In addition to the traditional classification based on duration, GRBs can also be categorized based on their prompt emission luminosities, such as high ($\ge$ 10$^{51}$ erg s$^{-1}$), intermediate ($\sim$ 10$^{49}$ - 10$^{51}$ erg s$^{-1}$), and low ($\leq$ 10$^{49}$ erg s$^{-1}$) luminosity GRBs. High-luminosity GRBs have been extensively studied, while low and intermediate-luminosity GRBs have emerged as an intriguing and less-explored domain \citep{2007ApJ...662.1111L, 2015PhR...561....1K}. 

\indent Following the initial burst, i.e., the spiky prompt emission phase, there is a steadily diminishing broadband afterglow phase originating from external shocks produced in the surrounding medium due to the interaction of the relativistic fireball with the circumburst medium \citep{2015PhR...561....1K}. In the pre \swift era, afterglow light curves across all energies typically decayed following power laws with one or more breaks, as predicted by various physical mechanisms, such as synchrotron emission from forward shocks in the external medium \citep{1999ApJ...520..641S}. In the \swift era, the early afterglow light curves sometimes exhibit features such as flares (mainly due to internal origin), bumps (mainly due to external origin), and plateaus, which are not explained by the external shock model and need additional components \citep{2010ApJ...720.1513K, 2023Univ....9..113O}. Flares, indicative of late central engine activity, commonly appear in both X-ray and optical light curves \citep{2005Sci...309.1833B, 2013ApJ...774....2S}. However, bumps are more frequently observed in optical/NIR light curves than in X-ray ones \citep{2010ApJ...725.2209L}. According to several studies \citep{1999ApJ...520..641S, 2007A&A...469L..13M}, when the relativistic fireball interacts with the surrounding medium, it begins to accumulate material from that medium. Initially, the energy of the fireball remains nearly constant until it has accumulated a sufficient amount of material so that the rest mass energy equals the initial kinetic energy corresponding to the bulk Lorentz factor ($\Gamma_0$). At this point, the light curve reaches its peak time ($t_{\rm p}$). The corresponding distance at this point is defined as the deceleration radius ($R_{\rm dec}$) and is seen as a peak in the afterglow light curve at a time $t_{p}$ \citep{2013NewAR..57..141G}.
Following this, the fireball enters a self-similar phase, and the observed afterglow emission begins to wane following a power law, as predicted by the external forward shock model \citep{1976PhFl...19.1130B}. \citealt{2010ApJ...725.2209L} analyzed the onset features in 17 optical and 12 X-ray afterglow light curves. They also investigated correlations among the parameters derived from the onset bump, finding that many of these parameters exhibited significant correlations with each other.

\noindent The early bump could also be due to the reverse shock that travels back into the ejecta when the fireball interacts with the surrounding medium \citep{2005ApJ...628..315Z}. Reverse shocks can occur in the thick shell or thin shell regime, depending on whether the shock travel time is shorter or greater than the prompt emission duration \citep{2015AdAst2015E..13G}. The presence of a reverse shock can be seen as a bump in the early afterglow light curve and can be helpful in constraining the magnetic properties of the fireball. Therefore, early observations of GRB afterglows are crucial to witness such features \citep{2021MNRAS.505.4086G}. The launch of the \swift satellite in 2004 revolutionized early afterglow observations by providing the precise localization of bursts within a few arcminutes, allowing for ground-based robotic optical telescopes to obtain the earliest possible observations for many GRBs. Nevertheless, there have been a few instances (e.g., GRB 080319B) where wide-field cameras have coincidentally captured the prompt emission within their fields of view before \swift even slewed to the target \citep{2008Natur.455..183R}.

\noindent As discussed above, afterglow light curves from external forward shocks are predicted to follow a power-law decay with an index of approximately 1. However, at late times ($>$ 10$^{4}$ s), the emission often begins to exhibit a sharp decline with an index ($\alpha$) of about 2, generally characterized by a sharp break in the light curve known as a ``jet break." This jet break provides crucial information about the geometry and dynamics of the GRB event. A statistical study of 138 GRBs with observed jet break features was conducted by \citealt{2020ApJ...900..112Z}. Although the concept of a jet break in afterglow light curves is well-established for bright GRBs, its manifestations and implications in the context of low and intermediate-luminosity events remain relatively unexplored. Understanding jet breaks in these GRBs not only gives insights into the nature of the collimated relativistic outflows responsible for the observed gamma-ray emission but also provides essential insights into the central engines and the interaction between GRBs and their surrounding environments.

\indent GRB 210210A, detected by the \swift mission, is an intermediate luminosity GRB. The X-ray afterglow of this burst exhibited a rare early bump and a late-time break. These features motivated us for a detailed exploration of this event. This paper aims to provide a comprehensive description of the properties and origin of GRB 210210A. In Section \ref{sec:obs}, we give the prompt and afterglow observations of GRB 210210A. In Section \ref{results}, the results derived from analyzing the observed data are given. In Section \ref{disc}, we present our discussion. Finally, Section \ref{summary} presents a summary and conclusion of our work.

\section{Observations, data analysis, and Results} 
\label{sec:obs}

\subsection{Prompt observations and data analysis} 
\label{sec:prompt}
On 10 February 2021 at 02:00:27 UT (hereafter referred to as \tzero), The Burst Alert Telescope (BAT, \citealt{2005SSRv..120..143B}) instrument onboard \swift first detected \thisgrb. Due to its rapid slewing capabilities, \swift's X-ray Telescope (XRT) and Ultra-Violet and Optical telescope (UVOT) instruments swiftly localized the burst within a 1.1 $\arcsec$ error box. \thisgrb was also detected by \kw during its prompt emission phase. The preliminary analysis of the \kw time-averaged spectrum from \tzero-1.685 s to \tzero+7.147 s, covering an energy range of 15–1500 keV, indicates that the best fit is achieved with a cutoff power law (CPL) with a power law index $\Gamma_{CPL}$ = -1.68$_{-0.23}^{+0.25}$ and peak energy \Ep = 16.6$_{-10.7}^{+7.2}$ \keV, indicating that \thisgrb should be classified as a very soft burst \citep{2021GCN.29517....1F}.

\subsubsection{Temporal and Spectral analysis of BAT data}
\begin{figure}[!ht]
\centering
\includegraphics[width=\columnwidth]{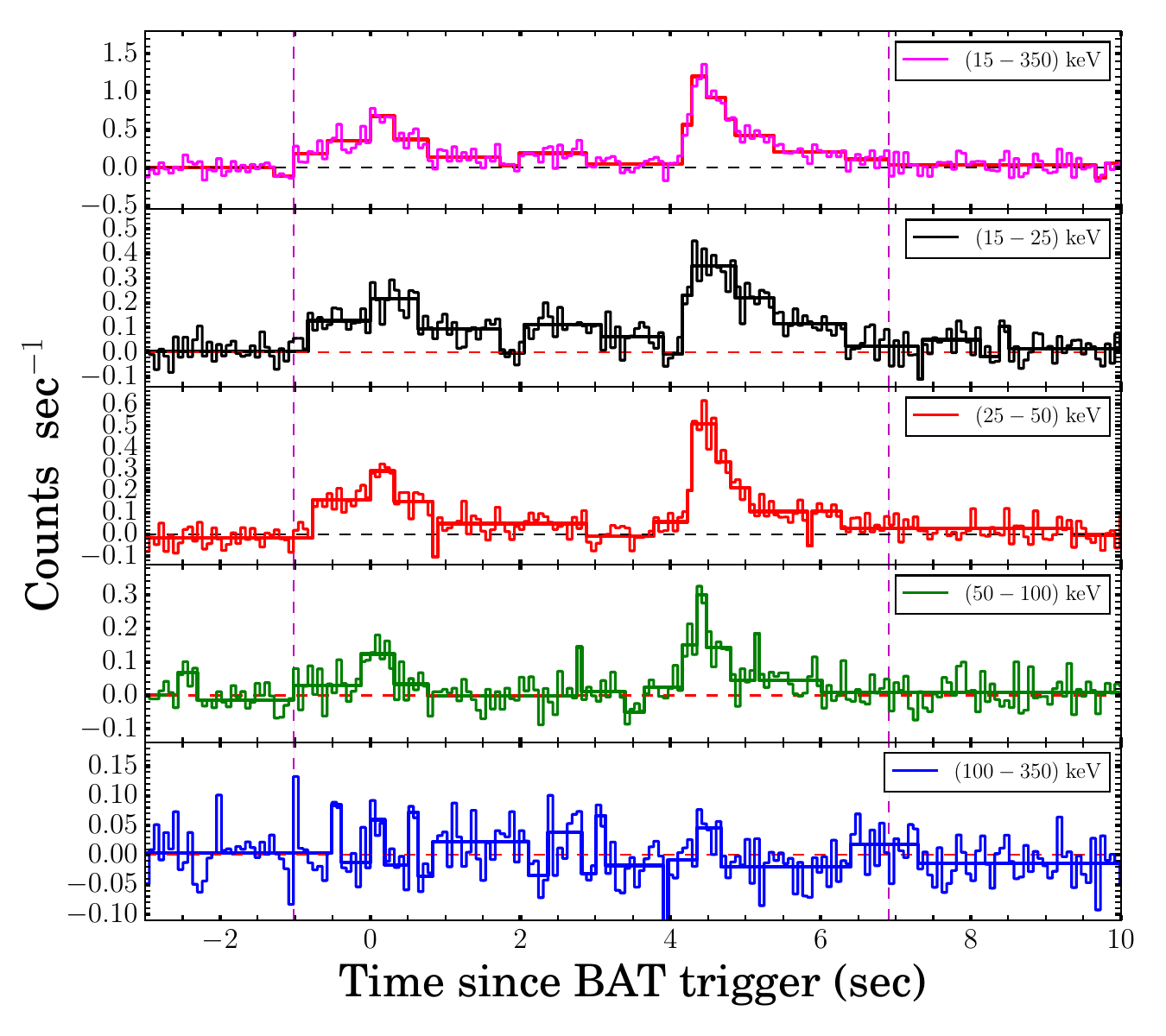}
\caption{\swift BAT prompt emission multi-channel mask-weighted light curves (temporal bin size of 64 ms) of GRB 210210A along with corresponding Bayesian blocks. The panels from top to bottom represent the light curve in the energy ranges [15–350], [15–25], [25–50], [50–100], and [100–350] \keV, respectively. Horizontal lines in each panel illustrate the background level, and the two vertical magenta lines cover the total duration used for the time-averaged spectrum.}
\label{fig:BAT_LC}
\end{figure}

We have used the methods discussed in \citet{2021MNRAS.505.4086G} to download and analyze the BAT observations. The prompt emission multi-wavelength light curve of \thisgrb is depicted in Fig. \ref{fig:BAT_LC}. The prompt emission's light curve consists of two distinct episodes; the first episode peak around at \tzero s, and the second episode peak around 4.2 s post-trigger, with a quiescent phase of 2 seconds. Most of the energy budget of this burst is concentrated around the second episode. The observed light curve of GRB 210210A is comparable to that of the prompt emission light curve of another well-studied low luminosity GRB 190829A but has a much shorter \tninty duration (6.60 $\pm$ 0.59 \,s, \citealt{2021GCN.29467....1L}).

We modeled the observed \swift-BAT spectrum in the 15–150 keV energy range. For this spectral fitting, we used a Python-based package the multi-mission maximum likelihood framework (\sw{3ML}), developed by \citet{2015arXiv150708343V}. We applied the Bayesian analysis method to fit the spectrum using various built-in models available in \sw{3ML}. We employed a multinest sampler, conducting 10,000 iterations to explore the parameter space thoroughly. To compare the effectiveness of the different fitted models, we utilized the deviance information criterion (DIC). We determined that a cutoff power-law model provides the best fit of the observed BAT spectrum (from \tzero-1.02 s to \tzero+6.91 s). The resulting parameters are as follows: the spectral index, $\Gamma_{CPL}$, is -1.34$_{-0.26}^{+0.26}$, and the spectral peak energy, \Ep, is 22.54$_{-12.54}^{+12.54}$ keV. These findings are compatible with those notified by \citet{2021GCN.29467....1L}, and \citet{2021GCN.29517....1F}. From this spectral analysis, we derived a fluence (in the 15-150 keV range) of 1.04 $\times$ 10$^{-6}$ erg cm$^{-2}$.

\subsection{Afterglow observations and data analysis} \label{sec:afterglow}

\subsubsection{X-ray afterglow analysis}

\begin{figure*}[!ht]
\centering
\includegraphics[width=0.69\columnwidth]{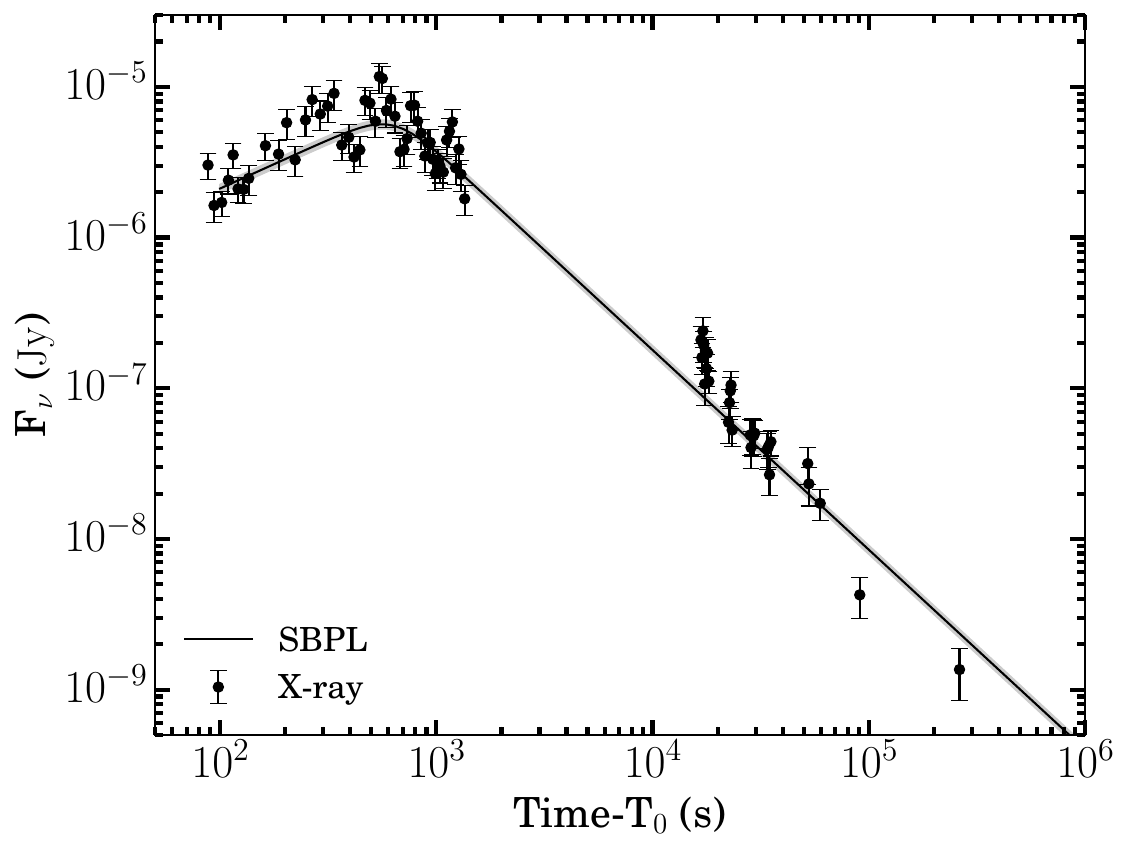}
\includegraphics[width=0.69\columnwidth]{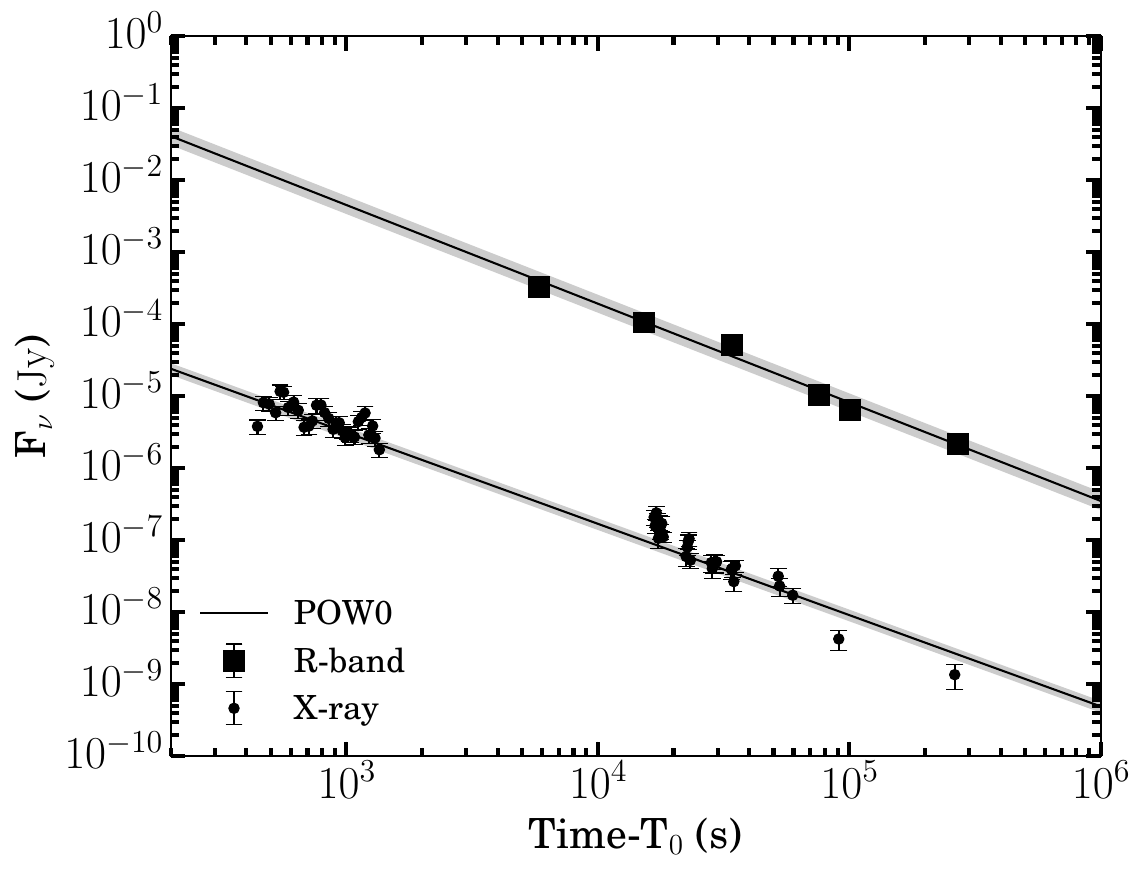}
\includegraphics[width=0.69\columnwidth]{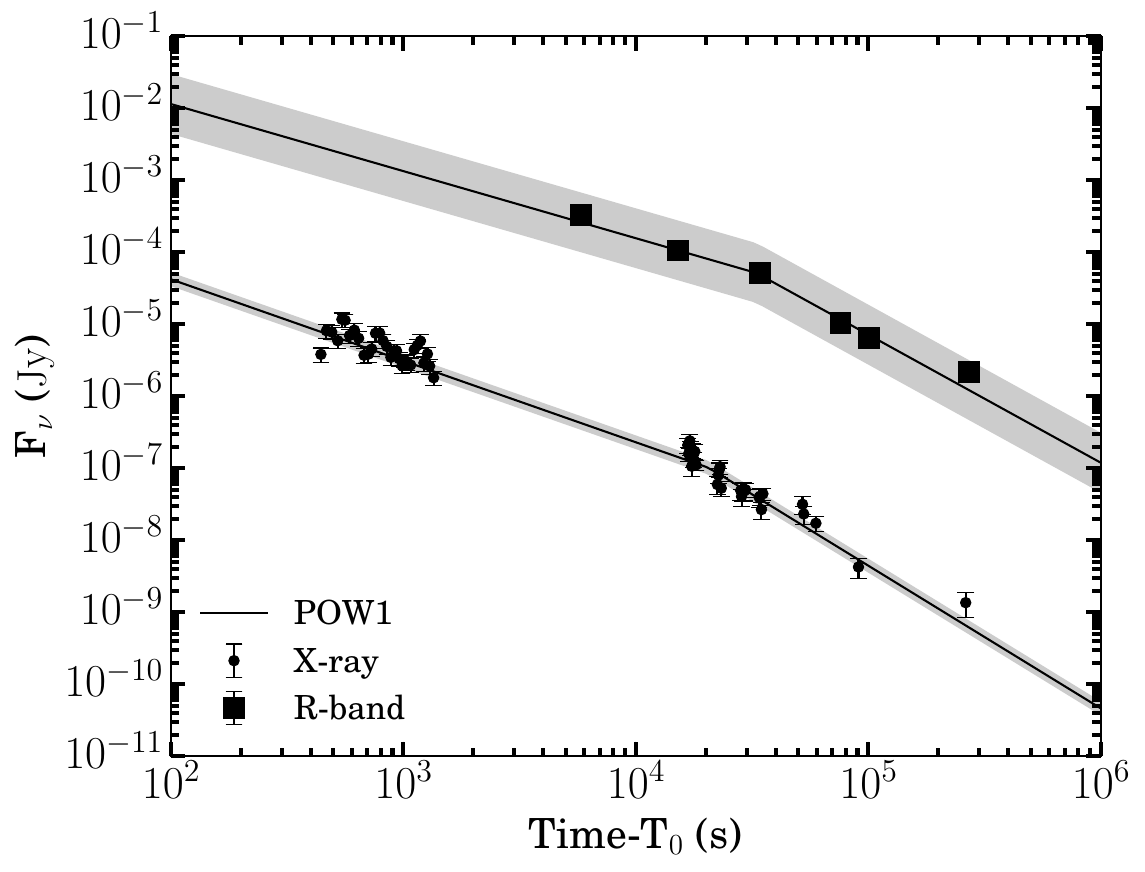}
\caption{Left panel: the unabsorbed \swift-XRT flux density light curve (at 10 keV) for GRB 210210A. Here, the circles indicate the \swift-XRT observations and the solid line shows the best-fit model using a smoothly joined broken power law. Middle panel: The same XRT light curve is now fitted with a simple power law after the initial bump was removed for clarity. The circles depict the XRT data, while the squares represent the optical R-band light curves. Right panel: Same as the middle panel but fitted with a single break power-law function.}
\label{fig:xlc}
\end{figure*}

Afterglow follow-up observations for the burst began promptly after \swift-XRT identified the potential X-ray counterpart of \thisgrb at \tzero + 82.1 s. Initially, \swift-XRT observed GRB 210210A in window timing mode for approximately 50 seconds before quickly transitioning to photon counting mode based on the low intensity of the burst. We accessed the \swift-XRT data from the \swift online repository\footnote{\url{https://www.swift.ac.uk/burst_analyser/01031728/}} \citep{2007A&A...469..379E, 2009MNRAS.397.1177E}. The observed \swift-XRT light curve revealed an early bump at approx. 600\,s after the trigger, followed by a power law decay segment. We fitted the unabsorbed \swift-XRT flux density light curve (at 10 keV) using a smoothly joined broken power law and employed the MCMC technique with 10,000 iteration steps for the fitting, discarding the first 500 as burn-in. The best-fit parameters were a rising index $\alpha_{r}$ = -0.65$_{-0.07}^{+0.07}$, a break time t$_{b}$ = 630.79$_{-39.82}^{+45.85}$ s, and a decay index $\alpha_{d}$ = 1.32$_{-0.02}^{+0.02}$. The obtained reduced chi-square value, $\chi^{2}_{\nu}$ = 2.3, was high, potentially due to deviations in the last two data points from the fit, which might indicate a late time break. Additionally, the XRT count rate light curve (in 0.3-10 keV) could also be successfully fitted with a broken power-law, indicating a plateau phase with parameters $\alpha_{1}$ = 0.31$_{-0.05}^{+0.05}$, a break time t$_{b}$ = 4529$_{-883}^{+982}$ s, and $\alpha_{2}$ = 1.72$_{-0.13}^{+0.15}$. To investigate the possibility of a jet break in the X-ray light curve, we re-analyzed the XRT flux density light curve after excluding the data points from the initial rise. Initially, we applied a simple power law fit, which yielded a decay index of $\alpha$ = 1.27$_{-0.02}^{+0.01}$ and a $\chi^{2}_{\nu}$ of 2.83. The elevated $\chi^{2}_{\nu}$ value, influenced by deviations in the last two points, suggested the presence of an additional break. Subsequently, we fitted the light curve with a broken power law, achieving as optimal fit parameters: a decay index before the break of $\alpha_{1}$ = 1.13$_{-0.08}^{+0.09}$, a break time t$_{b}$ = 10000$_{-1464}^{+2644}$ s, and a decay index after the break $\alpha_{2}$ = 1.98$_{-0.11}^{+0.12}$, with a $\chi^{2}_{\nu}$ of 1.92. The improvement in $\chi^{2}_{\nu}$ indicates an additional break around 10000s, likely representing a jet break in the X-ray light curve. The \swift-XRT light curve, along with the various power law models fitted to it, are illustrated in Fig. \ref{fig:xlc}. For a uniform external medium and an electron distribution with an index of $p=2.2$, the expected power-law indices before and after a jet break are $\alpha_1\sim0.9$ and $\alpha_2\sim p$. For a wind medium, the expected values are $\alpha_1\sim1.4$ and $\alpha_2\sim 1.65$.

\subsubsection{UV and Optical afterglow analysis}

The earliest optical observations available for the burst were taken by \swift-UVOT in the white filter at around \tzero + 85\,s. The UVOT data were analyzed using the standard HEASOFT UVOT tools (e.g., {\tt uvotproduct} and {\tt uvotmaghist}), see more details in \citet{2020ApJ...898...42C}, and \citet{2022MNRAS.511.1694G}. \thisgrb was successfully detected across all seven UVOT filters \citep{2021GCN.29457....1B}, and the results of this photometry are detailed in Table \ref{tab:UVOT} (appendix). No corrections to the UVOT data have been made for the expected extinction in the Milky Way corresponding to a reddening of $E_{\rm B-V}$ of 0.097\,mag.\ in the direction of the GRB \citep{1998ApJ...500..525S}, or for the GRB's host galaxy extinction. 

In addition to \swift-UVOT, \thisgrb was observed by multiple ground-based observatories, including the 1.3\,m Devsthal Fast Optical Telescope (DFOT) located at Devasthal, India \citep{2023arXiv231216265G, 2022JApA...43...82G, 2023arXiv230715585G}. The redshift of the burst, determined to be $z$ = 0.715, was identified through spectroscopic observations made by the 10.4\,m GTC \citep{2021GCN.29450....1D}. We monitored the optical afterglow of \thisgrb using DFOT on 2021-02-13 at 23:20:16 UT, which is 3.89 days after the \swift-BAT trigger. During these observations, we observed fifteen frames using the R filter, each with an exposure time of 180 seconds. We followed standard IRAF \citep{1993ASPC...52..173T} procedures for image cleaning (e.g., bias subtraction, flat field correction, and cosmic ray removal). Subsequent PSF photometry was performed using DAOPHOT \citep{1987PASP...99..191S}. However, the optical afterglow was not detected in the final stacked image, with an upper magnitude limit of 21.8. 

The observed optical light curve from most of the \swift-UVOT filters (v, b, u, uvw1, uvm2, uvw2) exhibit a decay consistent with a simple power law, characterized by the decay indices given in Table \ref{tab:model}. However, the white band UVOT light curve shows a significant deviation from a simple power law. A broken power law is found to be the best to describe the light curve in the white band, as shown in Fig. \ref{fig:r_x_SED} and Table \ref{tab:model}. For our analysis, we focused on the R-band light curve, which has a sufficient number of data points after \tzero+10000 s. Initially, we modeled the observed R-band light curve with a simple power law, obtaining a decay index of $\alpha = 1.37_{-0.03}^{+0.03}$, with a $\chi^{2}_{\nu}$/DOF = 13.5/4. Subsequently, we applied a broken power law fit to the same light curve and determined the decay indices to be $\alpha_{1} = 0.93_{-0.10}^{+0.08}$ before the break and $\alpha_{2} = 1.78_{-0.08}^{+0.09}$ after the break time $t_{b}$ = 30000$_{-3000}^{+2500}$ s, with a $\chi^{2}_{\nu}$/DOF = 9.26/2. Although the $\chi^{2}_{\nu}$ values did not approach unity in either fit, which may be attributed to the limited number of data points and the small error bars of the observations, the detection of a break in the optical light curve is statistically significant. The fitted models and the observed R-band light curve are presented in Fig. \ref{fig:xlc}.

\section{Results} \label{results}

\subsection{Prompt properties of \thisgrb}
We discuss the key prompt behavior of \thisgrb in the context of the larger sample.

\subsubsection{Classical and machine learning-based classification of \thisgrb} 

\begin{figure}[!ht]
\centering
\includegraphics[width=\columnwidth]{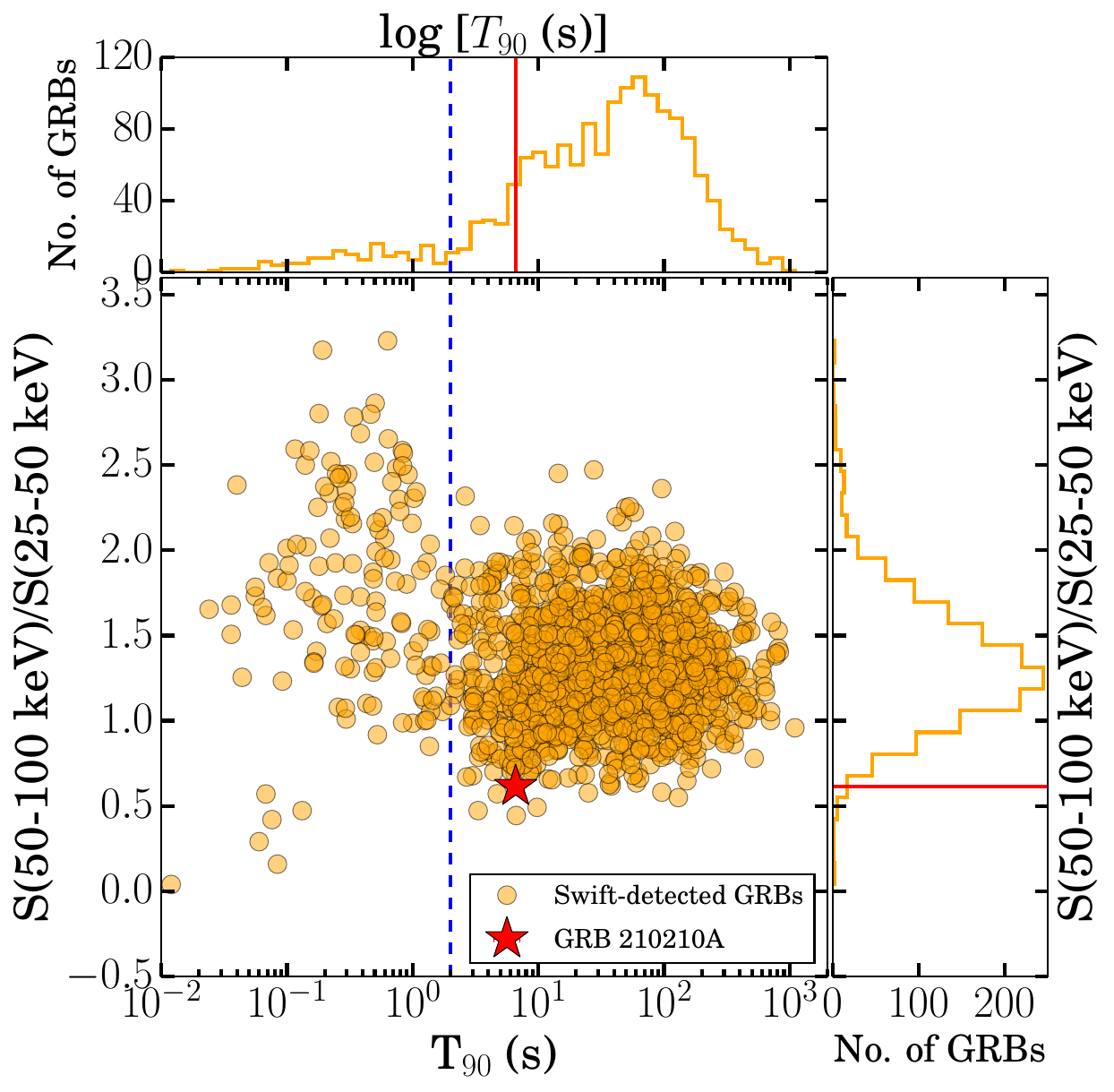}
\includegraphics[width=\columnwidth]{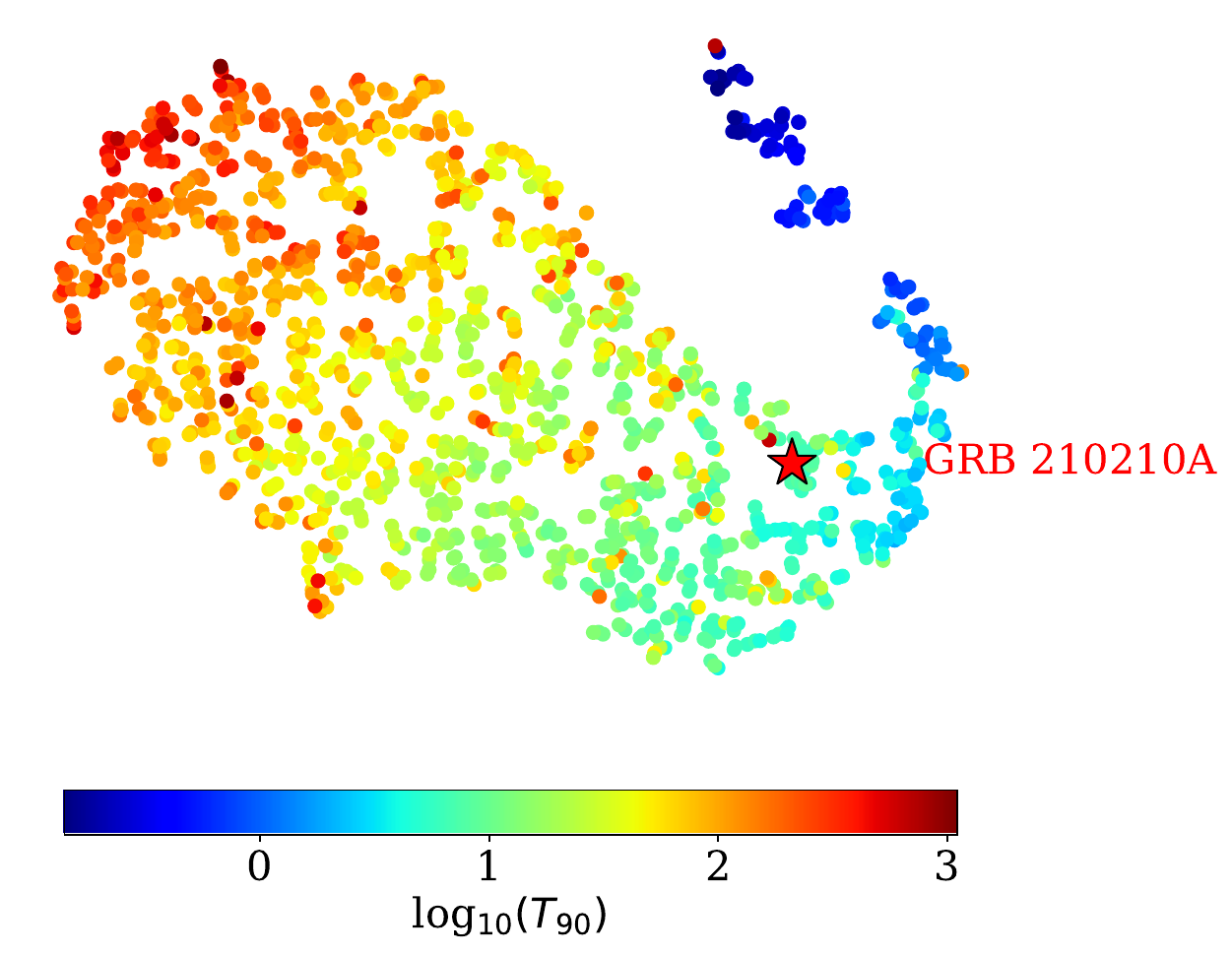}
\caption{Classical and machine learning-based classification of \thisgrb: Top panel represents \thisgrb in hardness ration (HR) - \tninty (dashed blue line is the boundary of classical classification) plane. The corresponding histograms of hardness ratio and \tninty (\thisgrb is highlighted with a solid red line) are also shown. {Bottom panel:} T-SNE distribution of \swift-detected GRBs. GRB 210210A is shown with a red star lying among the population of LGRBs.}
\label{fig:classification}
\end{figure}

The classification of GRBs has traditionally been based on their duration and spectral hardness. One commonly used method employs the \tninty duration—the time during which 90\% of the burst's flux is observed, from 5\% to 95\% of the cumulative flux in the 50 - 300 \keV band. There is an observed bimodality in the distribution of \tninty measurements with a separation at $\sim 2$\,s, leading to the separation of short (\tninty $\leq$ 2\,s) 
and long GRBs (\tninty $>$ 2\,s). However, this parameter has significant overlap, and \thisgrb lies in this \tninty overlapping region. We further obtained the spectral hardness ratio (HR) for \thisgrb equal to 0.61 by comparing the fluence values in two different energy bands (50-100 keV/ 25-50 keV). Our analysis helps to classify GRBs and suggests that \thisgrb is among one of the softest long bursts ever observed using \swift BAT (see Fig. \ref{fig:classification}).\\

{Classification using t-SNE:}
The t-Distributed Stochastic Neighbor Embedding (t-SNE) is a powerful tool for data visualization that represents high-dimensional data points in a lower-dimensional space. t-SNE is exceptionally effective in capturing non-linear relationships among data points, which makes it highly valuable for visualizing complex datasets. In the case of GRBs, t-SNE has been used to compare the prompt emission light curve of GRBs. Based on similarities and dissimilarities between the observed light curves, it places them on a two-dimensional map \citep{2023ApJ...951....4G}. The t-SNE map of \swift detected GRBs is shown in Fig. \ref{fig:classification}. The axes of this map do not have any meaning. However, the effects of \tninty on the grouping of GRBs are shown with different colors. As indicated with a color bar, the GRBs on the right of the map are short GRBs with \tninty $<$ 2s and mostly lie in the tail of the map. On the other hand, the bulk portion of the map consists of long GRBs. However, there is no sharp boundary between the two groups. GRB 210210A lies in the separation region between two types of GRBs. However, the observed properties of the burst are more consistent with LGRBs \citep{2024arXiv240601220R}. 

\subsubsection{High-energy correlation: Amati and Yonetoku}

\begin{figure}[!ht]
\centering
\includegraphics[width=\columnwidth]{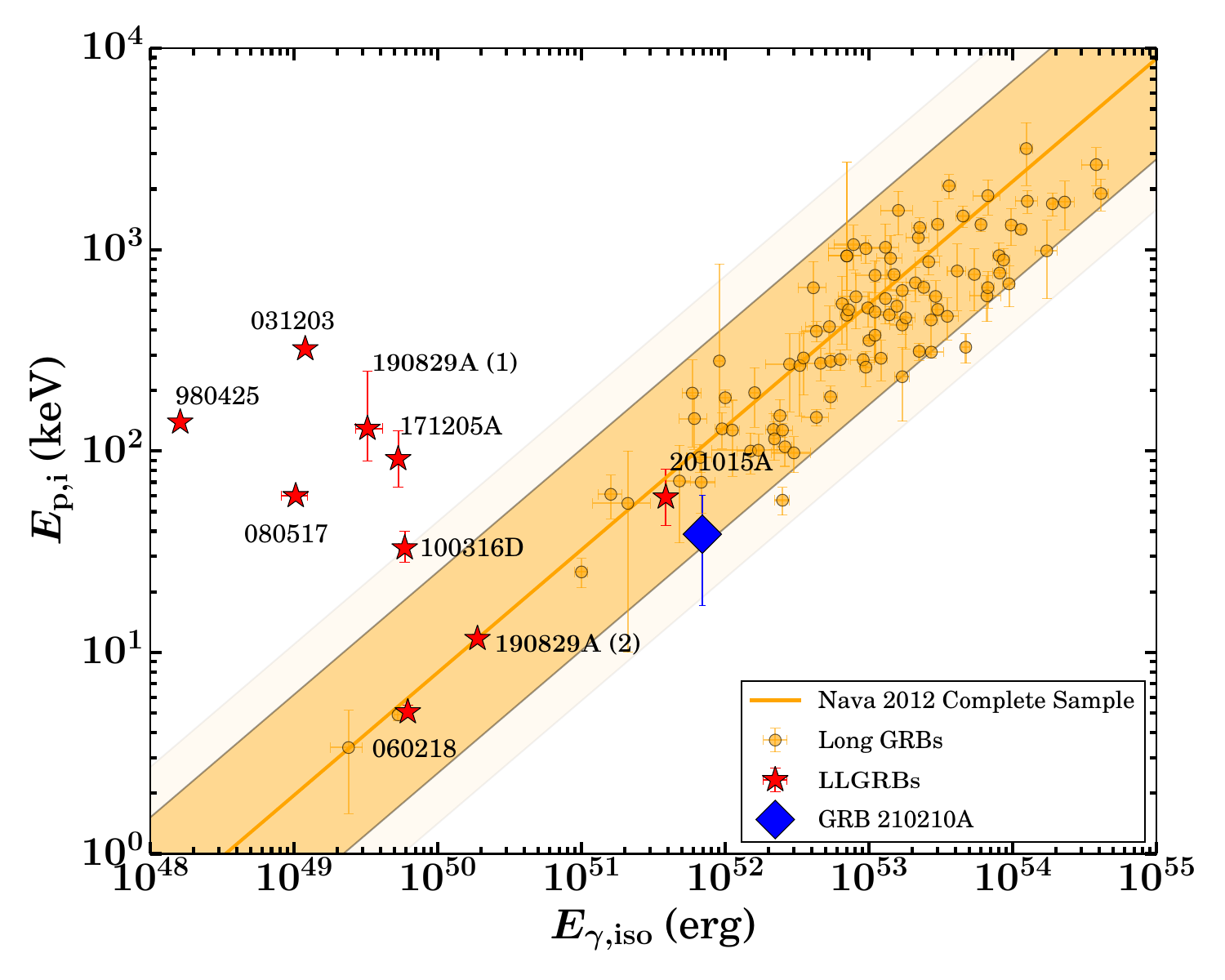}
\includegraphics[width=\columnwidth]{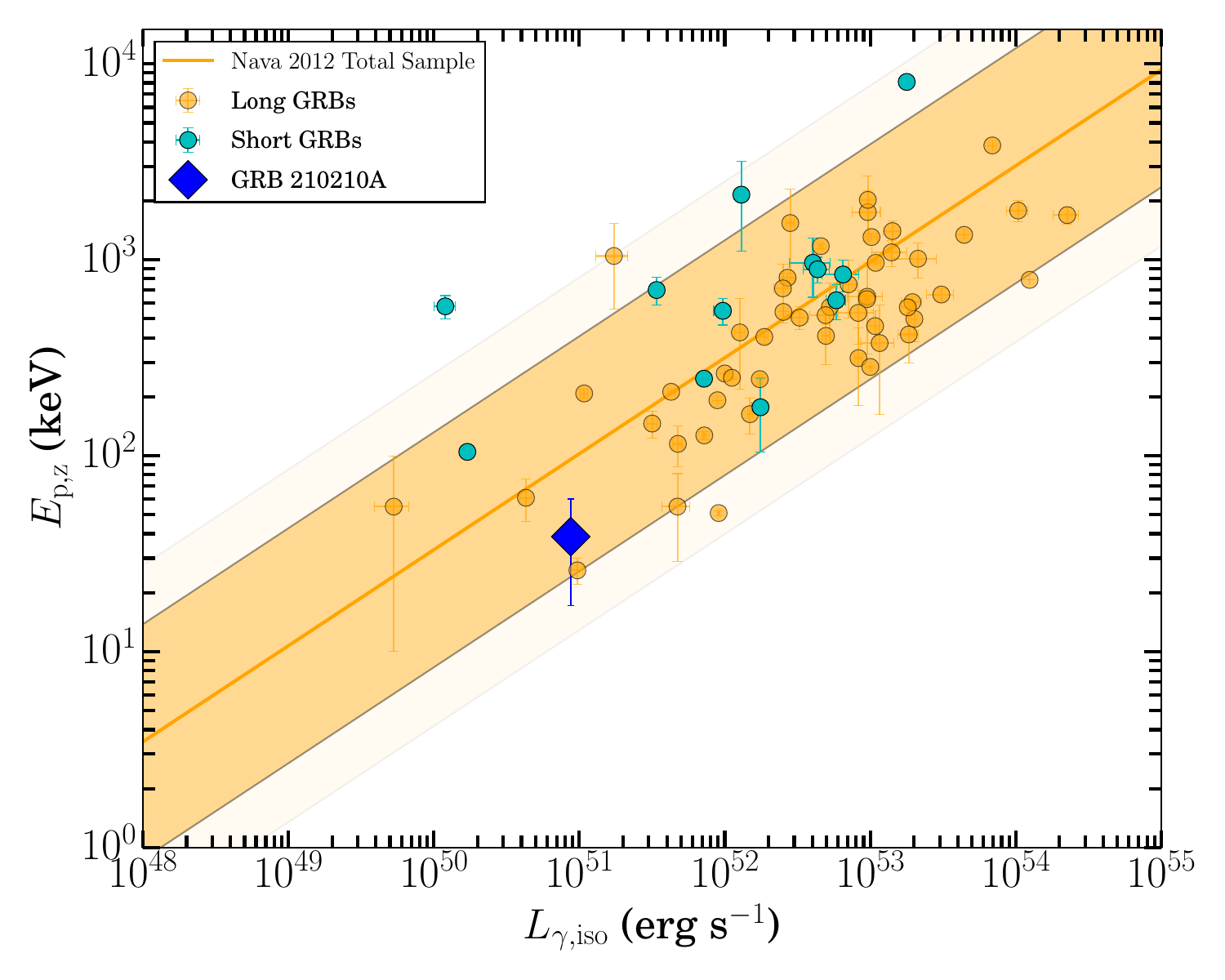}
\caption{Prompt emission correlations (\Ep-E$_{\rm \gamma, iso}$, Amati on top and \Ep-L$_{\rm \gamma, iso}$ Yonetoku on bottom) for GRBs along with \swift detected \thisgrb (shown with blue diamond). \thisgrb marginally satisfies the 2 $\sigma$ of the correlations but is consistent with 3 $\sigma$ of both correlations \protect\citep{2012MNRAS.421.1256N}. In Amati, we have also shown the positions of LLGRBs (with red stars) obtained from \protect\citet{2020ApJ...898...42C}.}
\label{fig:eiso_ep}
\end{figure}

The Amati relation suggests that there is a tight correlation between the isotropic energy (E$_{\rm \gamma, iso}$) emitted by a GRB and the peak energy of its gamma-ray spectrum in its rest frame \citep{2006MNRAS.372..233A}. In simple terms, brighter bursts tend to have higher spectral peak energies. To estimate the E$_{\rm \gamma, iso}$ for \thisgrb, we employed the method given in \citet{2015ApJ...815..102F}. With the observed spectral peak energy and the isotropic equivalent energy E$_{\rm \gamma, iso}$ = 6.92 $\times$ 10$^{51}$ erg (intermediate energetic), GRB 210210A marginally satisfies the 2 $\sigma$ of the correlation. Still, it is consistent with 3 $\sigma$ of the Amati correlation. The position of \thisgrb within the Amati correlation, along with other luminous and low-luminous GRBs, is presented in Fig. \ref{fig:eiso_ep}. 

On the other hand, the Yonetoku correlation relates the spectral peak energy (\Ep) of GRBs with the luminosity (L$_{\rm \gamma, iso}$) in their rest frame. Similar to Amati, with the observed spectral peak energy and the isotropic luminosity L$_{\rm \gamma, iso}$ = 8.73 $\times$ 10$^{50}$ erg s$^{-1}$ (intermediate luminosity), GRB 210210A marginally satisfies the 2 $\sigma$ of the correlation. Still, it is consistent with 3 $\sigma$ of the Yonetoku correlation (Fig. \ref{fig:eiso_ep}).

\subsection{Afterglow properties of \thisgrb}

In this subsection, by analyzing the afterglow emissions of \thisgrb, we aim to understand the dynamics of the burst, the nature of its immediate surroundings, and the mechanisms driving the emission processes. 

\subsubsection{Afterglow emission and spectral regime}

\begin{figure}[!ht]
\centering
\includegraphics[width=\columnwidth]{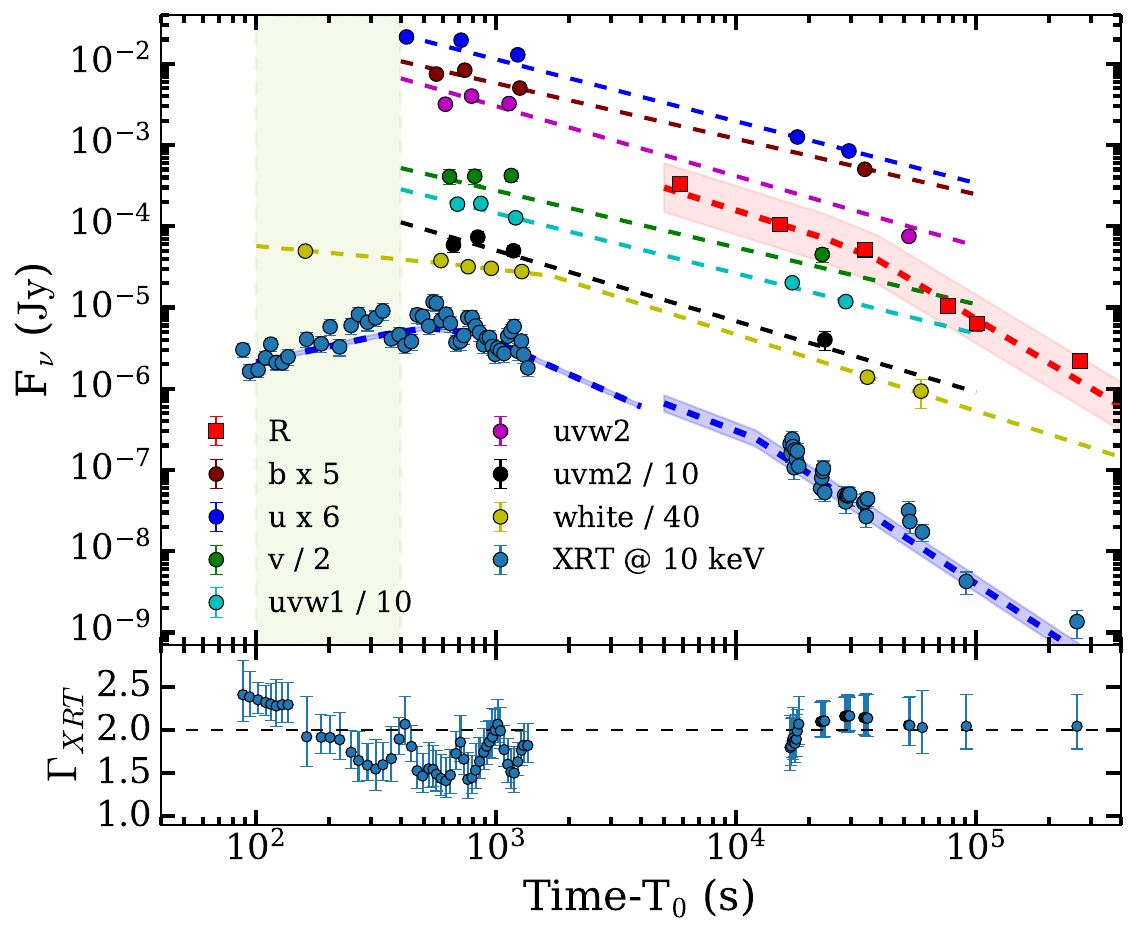}
\includegraphics[width=\columnwidth]{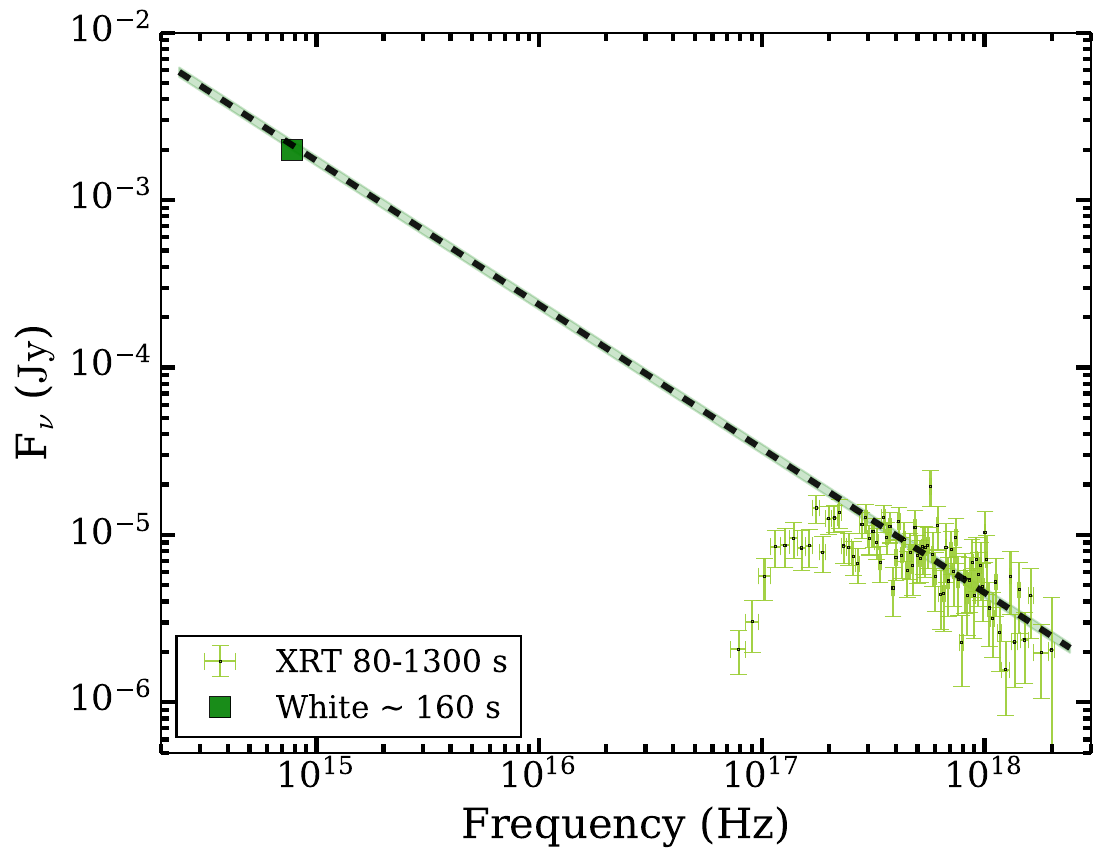}
\caption{Upper panel: The multi-wavelength afterglow light curve of GRB 210210A. The colored circles represent the \swift-UVOT observations fitted by the power law function, as shown with thin dashed lines. \swift-XRT observations, along with a smoothly joined broken power law fitted to it, are shown with a blue circle and thick blue line, respectively. The X-ray photon indices evolution in 0.3-10 \keV is shown at the bottom of the plot. The shaded region shows the X-ray and optical SED duration. Lower panel: optical-to-X-ray SED created at 100–1300\,s.}
\label{fig:r_x_SED}
\end{figure}

The X-ray and optical afterglow light curve of \thisgrb is plotted in Fig. \ref{fig:r_x_SED}. As the temporal analysis of early X-ray and optical emission shows a different behavior (possibly due to a few points during early optical observations), we explored the optical-to-X-ray spectral energy distribution (SED) at an early time (where the earliest UVOT observations are available). The observed XRT spectrum in 0.3-10\,keV was obtained from the \swift-XRT GRB spectrum repository \citep{2007A&A...469..379E, 2009MNRAS.397.1177E}. We fit the spectrum utilizing an absorbed power law model with Galactic (\sw{phabs}, N$_{\rm H, Gal}$) and host (\sw{zphabs}, N$_{\rm H, z}$) absorption components in \sw{XSPEC} software \citep{2022MNRAS.511.1694G, 2021RMxAC..53..113G}.  First, we used the value of N$_{\rm H, Gal}$ and constrained N$_{\rm H, z}$ by fitting the late-time XRT spectrum. By fixing the N$_{\rm H, Gal}$ = 8.82 $\times$ 10$^{20}$ cm$^{-2}$ and N$_{\rm H, z}$ = 2.35 $\times$ 10$^{21}$ cm$^{-2}$ from the late spectral fitting, we fit the early time X-ray spectra (in the temporal range 85-1350\,s and energy range 0.3-10 keV) and determine the observed spectral index is $\beta$ = 0.79$^{+0.09}_{-0.09}$. We extrapolated this index and found that the observed earliest optical (galactic extinction corrected) emission (in UVOT white filter) is consistent with the extrapolation of the X-ray spectral index. This indicates a negligible amount of host extinction if optical and X-ray emissions have similar origins (see Fig. \ref{fig:r_x_SED}). Further, our SED analysis suggests that the observed spectral and temporal indices correspond to the spectral regime $\nu_{o}$ $<$ $\nu_{x}$ $<$ $\nu_{c}$, in the ISM-like medium. We obtained the value of electron energy distribution index $p$ = $\frac{\beta-1}{2}$ = 2.6

\subsubsection{Late time possible jet break in \thisgrb}

The GRB emission is assumed to be produced by a relativistic jet. The evidence of the jet break is observed in the late time afterglow light curve as a steep decay phase ($\alpha \sim 2$) after a normal decay phase ($\alpha \sim 1$) around 10$^{4}$ - 10$^{5}$ s post-trigger. The observed jet break in the afterglow light curve is utilized to calculate the jet opening angle $\theta_j$ and related physical parameters. For GRB 210210A, we found that both X-ray and optical light curves have a statistically significant break around 1E+4\,s and 3E+4 \,s, respectively. The observed slope before and after the break time t$_{b}$ is $\alpha_{1}$ $\sim$ 1 and $\alpha_{2}$ $\sim$ 2 for both X-ray and optical light curves. This indicates that the observed break in the X-ray and optical light curve is possibly due to the jet break. Considering jet break is a geometric effect, we expect that there should not be a change in the spectral index before ($\beta_1$) and after ($\beta_2$) the break. We calculated $\beta_1$ and $\beta_2$ to be equal to 0.79$^{+0.09}_{-0.09}$ (large variation during this epoch) and 1.01$^{+0.13}_{-0.13}$, respectively, using XRT observations. We note that these values are consistent within the error bars, supporting that the jet break model. The jet break in the optical light curve of \thisgrb is also independently confirmed by \citet{2021GCN.29502....1K}. However, we noted that the observed jet break times in the X-ray and optical light curves are not consistent, showing a chromatic nature (although it may not be intrinsic, due to the less coverage in the optical band, we are not able to precisely constrain the break time). There are several possible reasons for this, such as a structured jet model or two jet components, energy injection, environmental effects, etc. Assuming a two-component jet model for \thisgrb (if the chromatic jet break is intrinsic), the X-ray emission comes from the narrow jet component, and the optical emission comes from the wide jet component. Utilizing the jet break time, we have calculated the jet opening $\theta_{j}$ utilizing the method/parameters given in \citet{2001ApJ...562L..55F}, and \citet{2022JApA...43...11G}. The calculated values of $\theta_{j}$ from the jet break time in X-ray (narrow jet) and the jet break time in optical (wide jet), respectively, are 2.63$^{\circ}$ and 3.97$^{\circ}$.

\section{Discussion} \label{disc}

\subsection{Origin of the observed early X-ray bump}

\swift-XRT observations revealed a bump in the X-ray light curve of \thisgrb. Since we do not have enough simultaneous optical/UV observations corresponding to the X-ray bump, it is hard to discuss any early achromatic behavior between the two bands. However, the earliest observations on the UVOT/white band are slightly inconsistent with a simple power law decay and may hint towards an achromatic bump occurring in both the optical and X-ray light curves. Nevertheless, we do not have enough optical observations to constrain the parameters of a bump (if any) in the early optical light curve. In the next sub-sections, we searched for the possible origin of this X-ray bump. We investigated if this bump could be the signature of a reverse shock, the onset of an external forward shock in the surrounding circumburst medium, or due to an observer viewing the burst slightly outside the narrow jet, i.e., $\theta_{obs}$ $>$ $\theta_{J}$, energy injection, etc \citep{2015PhR...561....1K, 2023Univ....9..113O}.

\subsubsection{Reverse shock}
During the interaction of the relativistic fireball with the surrounding medium, two types of external shock are generally assumed. A forward shock propagates into the surrounding medium and a reverse propagates shock back into the ejecta. Reverse shock emission is characterized by an early sharp rise in the afterglow light curve. However, reverse shock emission is not always visible if the outflow is strongly Poynting flux dominated \citep{2015AdAst2015E..13G}. In the case of reverse shock, the early optical/X-ray light curve in the ISM-like medium should rise by the index of -5 in the thin shell case, and -0.5 in the thick shell case, and the decay slope should be around 2 \citep{2015AdAst2015E..13G}. For the wind environment, the light curve should show an early rise of -2.5 and the decay slope around 3 \citep{2015AdAst2015E..13G}. For \thisgrb, the observed peak time t$_{\rm p}$= t$_{\rm b}$ $\times$ $\left(\frac{-\alpha_{r}}{\alpha_{d}}\right)^{1/w(\alpha_{d}-\alpha_{r})}$ = 559\,s is much greater than the observed \tninty duration of the burst, so it is reasonable to assume a thin shell case. The observed parameters of the bump $\alpha_{\rm r}$ = -0.65 and $\alpha_{\rm d}$ = 1.32 are not consistent with the above cases in both wind or ISM-like medium. Therefore, we discard the possibility of reverse shock emission dominating early X-ray bump observed in the case of \thisgrb.
 
\subsubsection{Onset of afterglow} 

\begin{figure}[!ht]
\centering
\includegraphics[width=\columnwidth]{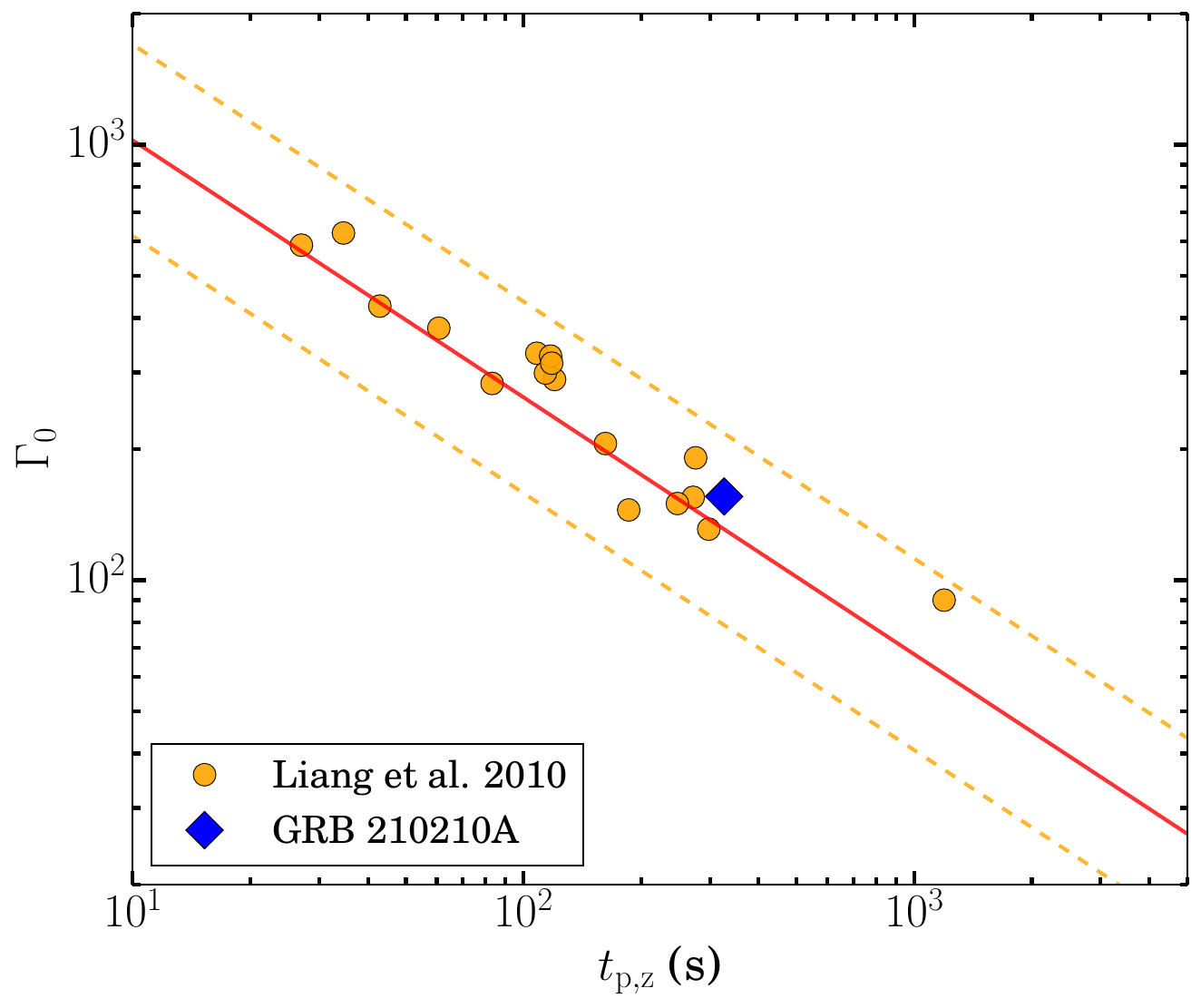}
\includegraphics[width=\columnwidth]{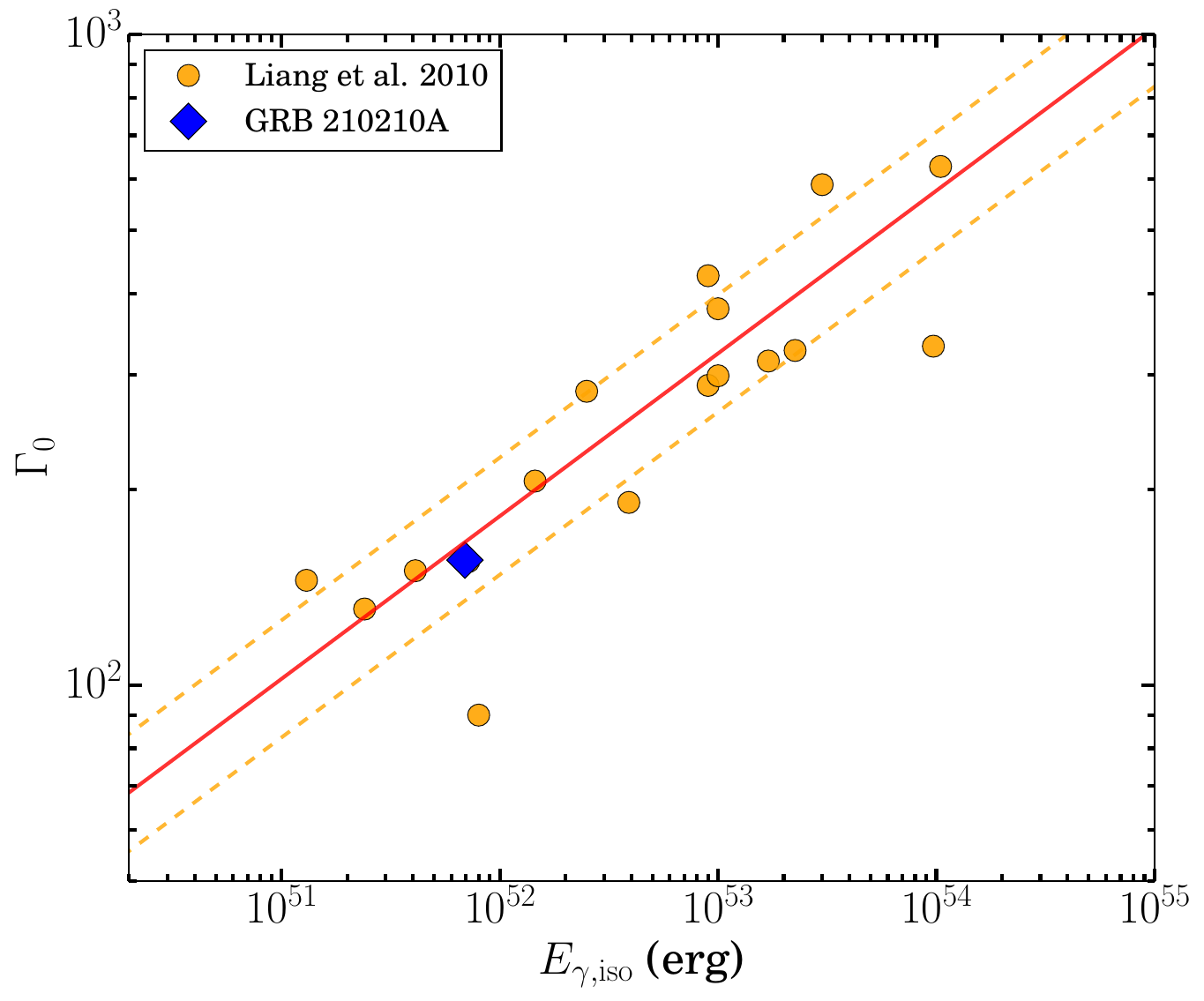}
\caption{The top panel represents the distribution of the Lorentz factor ($\Gamma_{0}$) along with the observed peak time (t$_{\rm p}$) of the onset bump in the rest frame. Solid and dotted lines represent the correlation found by \citet{2010ApJ...725.2209L} and corresponding 2$\sigma$ region. The lower panel similarly represents the distribution of the Lorentz factor and isotropic gamma-ray energy release (E${\rm \gamma, iso}$). GRB 210210A is shown with a blue diamond, and the orange circles are the data points taken from \citet{2010ApJ...725.2209L}.}
\label{fig:onset}
\end{figure}

\begin{figure}[!ht]
\centering
\includegraphics[width=\columnwidth]{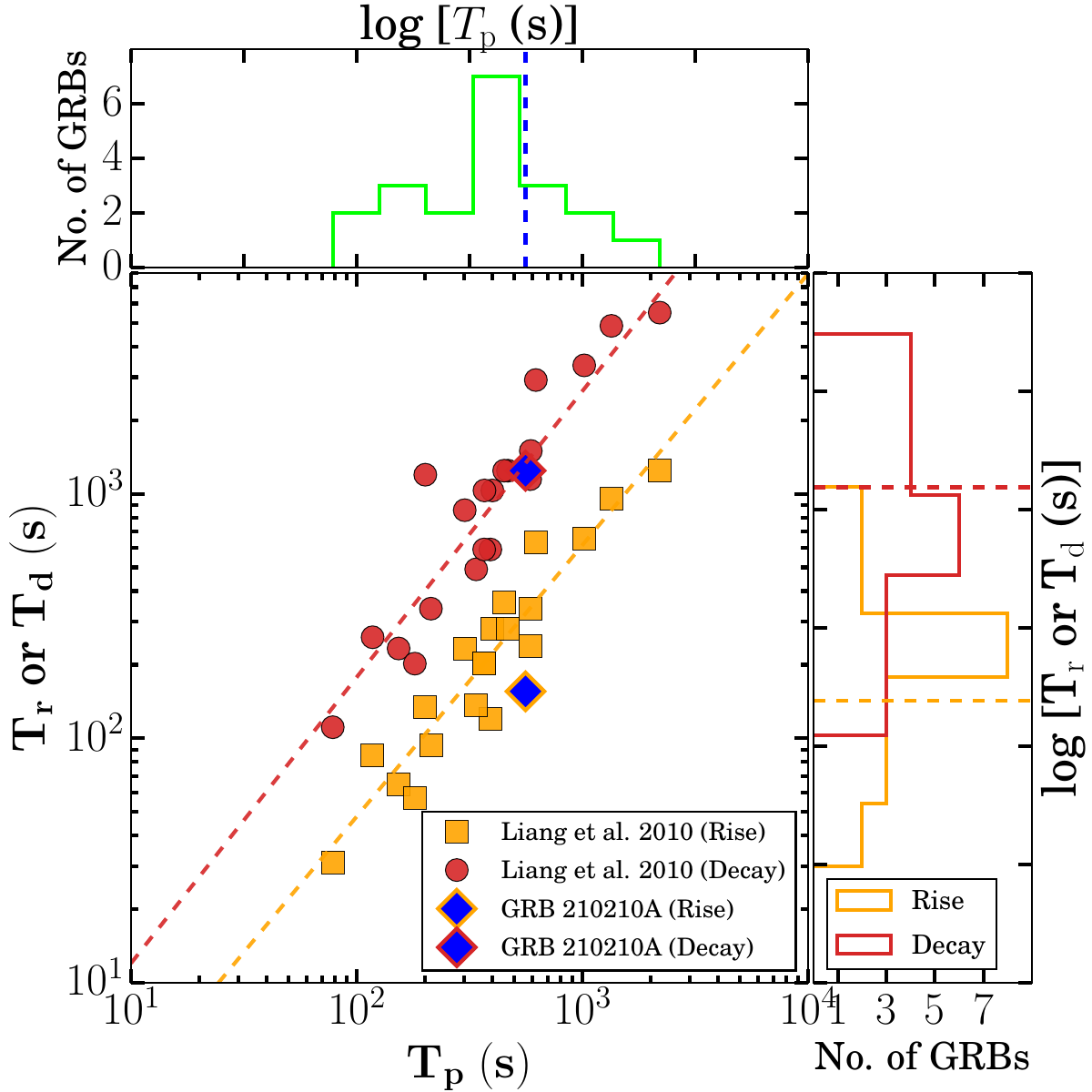}
\caption{The distribution of the rise and decay times with the observed peak time (t$_{p}$) of the onset bump. The red and orange dotted lines represent the correlation found by \citet{2010ApJ...725.2209L}. The histograms of the observed peak time (lime), rise time (orange), and decay time (red) are shown in the upper and right panels, respectively. The position of GRB 210210A is shown with vertical dashed lines in these histograms, and orange squares/red circles are the data points taken from \citet{2010ApJ...725.2209L}.}
\label{fig:onset2}
\end{figure}

\citealt{2010ApJ...725.2209L} extensively searched for the deceleration feature in the early light curves of X-ray and optical afterglows, and they found 20 optical (17 with redshift measurements) and 12 X-ray (only two with redshift measurements) afterglow light curves with onset features. The onset feature in optical light curves is common but rare in X-ray light curves. Additionally, they studied several correlations among the parameters of the onset bump obtained based on the temporal fitting of the bump using a smooth broken power-law function. To check the nature of the X-ray bump observed in our case, we fitted the bump using smoothly joined broken power law and calculated the bump parameters (see section \ref{sec:afterglow}). We utilized these parameters to calculate the bulk Lorentz factor ($\Gamma_{0}$ $\sim$ 156) from the relation given by \citet{2007A&A...469L..13M}. We placed these parameters in the correlations plane of \citet{2010ApJ...725.2209L} see also \citet{2023ApJ...942...34R}. In Fig. \ref{fig:onset}, we can see that the calculated value of the bulk Lorentz factor, t$_{p}$, and E$_{\rm \gamma, iso}$ for \thisgrb perfectly satisfies the observed correlation by \citet{2010ApJ...725.2209L}, supporting the onset origin for the observed early bump in the X-ray light curves. Additionally, we estimated the rise and decay time of \thisgrb and investigated the possible correlation of these parameters with peak time along with data points obtained from \citet{2010ApJ...725.2209L}. \thisgrb is consistent with these correlations (see Fig. \ref{fig:onset2}), further supporting the onset nature of the early X-ray afterglow bump.

\subsubsection{Off axis observation of \swift-XRT}
\label{offaxis}

\begin{figure}[!ht]
\centering
\includegraphics[width=\columnwidth]{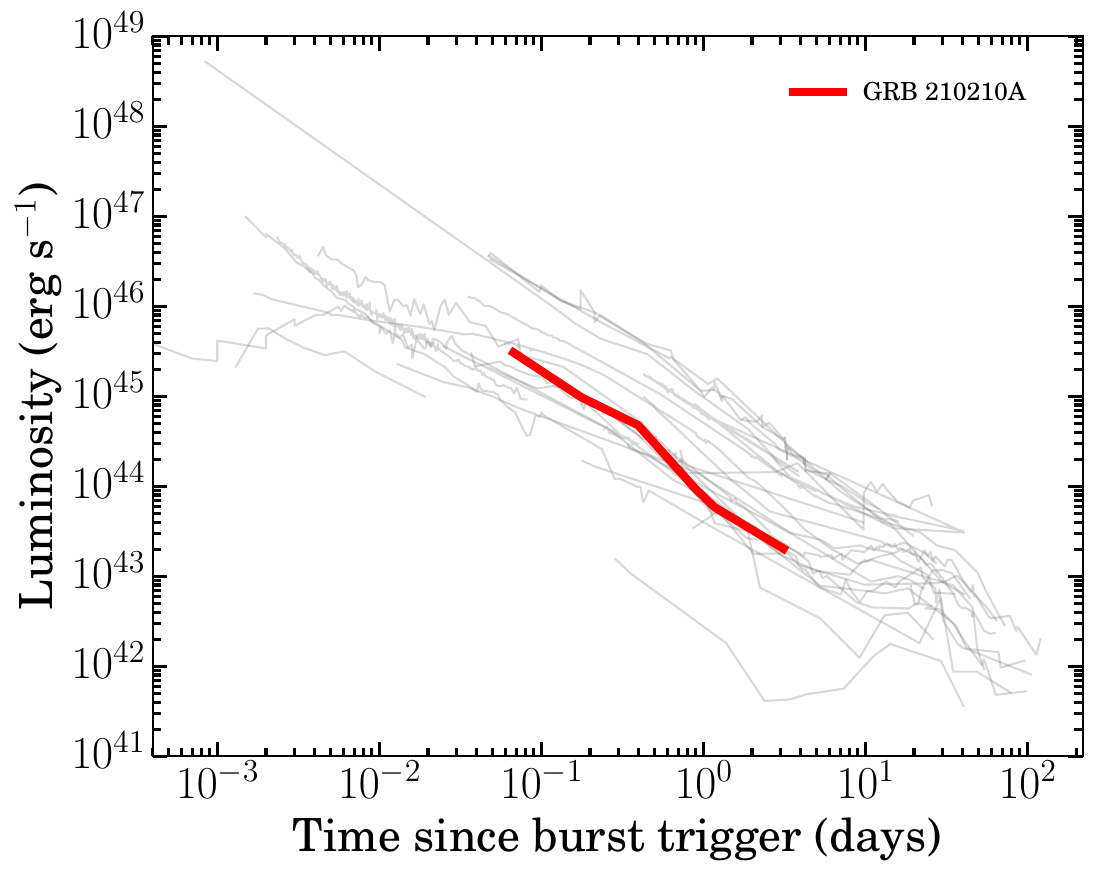}
\includegraphics[width=\columnwidth]{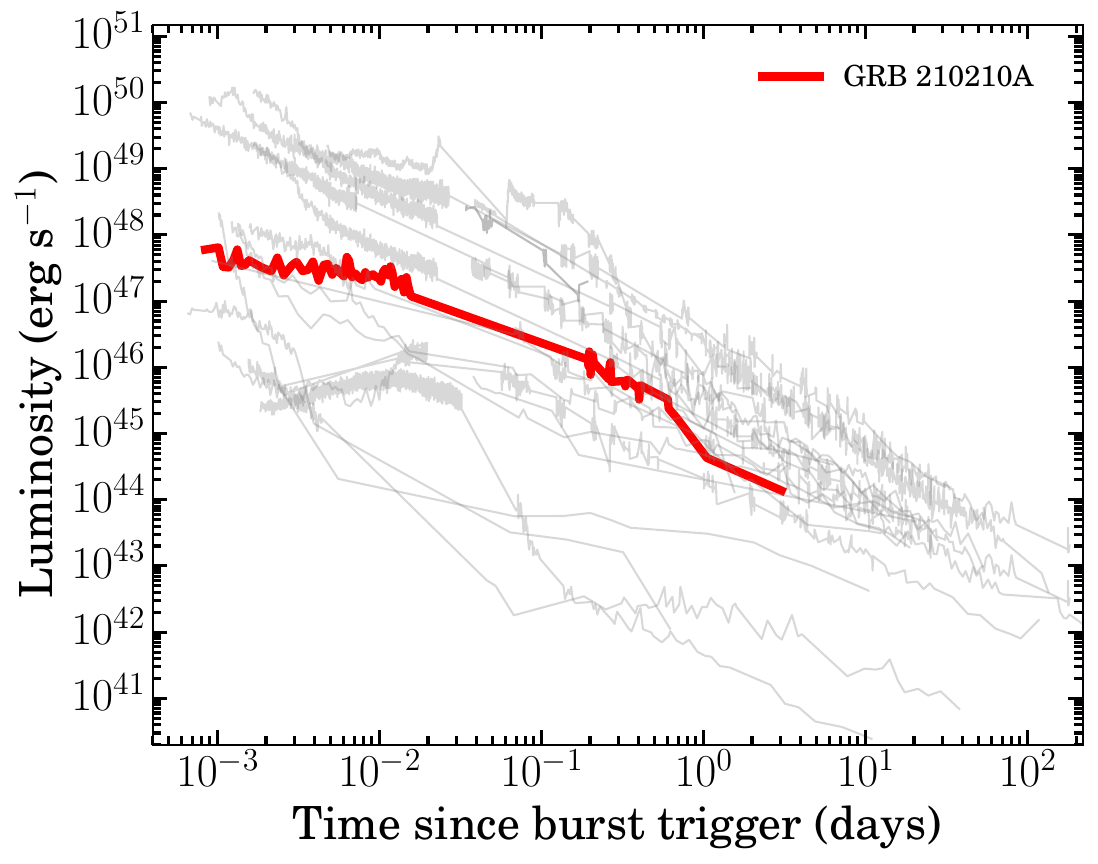}
\caption{Upper panel: the R-band optical luminosity afterglow light curve of \thisgrb (depicted by the red curve) compared with other nearby GRBs (shown in grey). Lower panel: the \swift-XRT X-ray luminosity afterglow light curve of \thisgrb (red curve) compared with other nearby GRBs (grey).}
\label{fig:r_x_cmp}
\end{figure}

One of the possible reasons for the early rise in the X-ray light curve could be if the observer is initially looking outside the jet (a GRB jet may be oriented away from Earth's line of sight), i.e., the jet opening angle is smaller than the viewing angle ($\theta_{j}$ $<$ $\theta_{obs}$). In such a case, when the jet loses its energy, it starts spreading, and the observer begins to observe more flux, and the light curve rises initially. We need detailed afterglow modeling to constrain the viewing angle. However, observational signatures of off-axis observations of X-ray afterglows exhibit distinct temporal and spectral features compared to on-axis observations, such as a typically lower X-ray flux (due to the emission being less beamed and hence appears fainter), a delayed onset of the afterglow (the emission becomes observable only after the jet has decelerated sufficiently for the relativistic beaming effect to expand the emitting region into the observer's line of sight), a shallower decay (the decay slope of the light curve post-peak will be shallower compared to on-axis observations due to the contribution of emission from wider angles becoming visible over time), an extended plateau phase (off-axis afterglows often show an extended plateau phase, where the brightness remains relatively constant for a longer period before declining). This is due to the gradual increase in the observed emitting area as the jet spreads laterally), reduced apparent luminosity (the observed luminosity of an off-axis GRB is lower because the relativistic beaming effect, which concentrates the jet's energy into a narrow cone, is less effective at larger viewing angles), etc. For \thisgrb, we investigated these features in its afterglow light curve. We noted that the X-ray flux observed at 11 hours and 24 hours post-burst is significantly brighter than the typical X-ray counterpart of long bursts \citep{2022JApA...43...11G}. An onset was observed in the early X-ray observations of \thisgrb, with no shallower decay emission, and no observed extended plateau phase plateau, indicating on-axis observations. We also compared its X-ray and optical afterglow luminosity with nearby GRBs and found that the initial X-ray and optical emission from \thisgrb has an intermediate luminous afterglow and that its late emission shows the jet break (see Fig. \ref{fig:r_x_cmp}). Additionally, we detected the jet break signature in the afterglow light curve of \thisgrb at the usually expected time from on-axis observations (the observation of a jet break may be less noticeable or occur at a later time in off-axis observations). This is due to the observer seeing emission from progressively wider angles as the jet slows and expands; further supporting an on-axis jet model for \thisgrb afterglow and discarding the off-axis scenarios.

\subsection{Origin of Intermediate luminosity of \thisgrb} 
Low-luminosity or intermediate-luminosity gamma-ray bursts can originate from various mechanisms and environments. In the present section, we discuss a few possible scenarios. 

\subsubsection{Shock Breakout from Stellar Explosions}
According to the shock breakout model, low-luminosity GRBs may result from the shock breakout of a supernova (SN) explosion (unsuccessful jets). Generally, these bursts are not consistent with the Amati correlation. When the shock wave from a supernova's core collapse reaches the surface of the star, it can produce a brief burst of gamma rays. In such cases, the duration of the burst follows a fundamental correlation
$T_{90}\sim 20~{\rm s}~(1+z)^{-1.68}\left(\frac{E_{\rm \gamma,iso}}{10^{46}~{\rm erg}}
\right)^{1/2}\left(\frac{E_{\rm p}}{50~{\rm keV}}\right)^{-2.68}$ \citep{2012ApJ...747...88N}. In the case of \thisgrb, we calculated the expected shock breakout duration to be approximately 56\,ks, which is much longer than the observed \tninty duration, thereby inconsistent with the predictions of the shock breakout scenario. Additionally, the estimated gamma-ray efficiency for \thisgrb is $\eta \sim$ 1\,\% (typical of long and short GRBs), inconsistent with those expected from shock breakout and LLGRBs ($\eta \sim 10^{-4}$\,\%, \citealp{2018MNRAS.479..588G}). Furthermore, the shock breakout model predicts a very simple light curve with a fast rise and exponential decay without any gap. However, for \thisgrb, we observed two episodic emissions with a significant gap, which further rules out the shock breakout model.

\subsubsection{Choked Jets}
For GRBs, a jet is a highly relativistic outflow of particles and radiation launched by the central engine, usually a collapsing massive star or a merger of compact objects. A choked jet occurs when this jet fails to escape the progenitor star before losing its energy. The jet is ``choked" because it does not break out of the stellar envelope, thus preventing a typical high-energy gamma-ray burst from being observed. In such a scenario, the jet's energy is dissipated into the stellar envelope, converting kinetic energy into thermal energy. This results in a sub-relativistic, mildly relativistic shock that can produce observable emissions (also possible neutrinos) different from standard GRBs \citep{2016PhRvD..93h3003S, 2018ApJ...855...37D}. Due to the shock heating of the stellar envelope, choked jets can produce thermal X-ray emissions that are detectable as a smooth thermal spectrum. Also, choked jets can produce strong radio emissions as the sub-relativistic shock-wave interacts with the surrounding interstellar medium. However, for \thisgrb, we do not detect any thermal signature in the X-ray regime, and no radio emission is reported, and we clearly detected the signature of jet break, thus discarding the choked jet scenario.

\subsubsection{Magnetar Central Engine?}

Two types of central engines are generally considered in GRB models. Either a black hole with an accretion disc that utilizes the rotational energy of the disc through the magnetic field \citep{1977MNRAS.179..433B} or a rapidly spinning magnetar (a neutron star with an extremely strong magnetic field) which provides energy through the spin-down of the magnetar. However, a magnetar central energy can only produce a GRB with a maximum energy of 10$^{52}$ erg; any GRBs crossing this must have a black hole central engine. A magnetar can also power a GRB with relatively lower luminosity. To determine the possible central engine of GRB 210210A, we applied the techniques described in \citet{2021ApJ...908L...2S}. The key idea of this approach revolves around the maximum potential rotational energy from a millisecond magnetar, which can launch jets with energies around 10$^{52}$\,erg. We calculated the beaming-corrected gamma-ray energy ($E_{\rm \gamma, beamed}$) for \thisgrb to be 7.30 $\times$ 10$^{48}$\,erg, which is within the possible energy budget of a millisecond mangetar central engine.

\begin{figure}[!ht]
\centering
\includegraphics[width=\columnwidth]{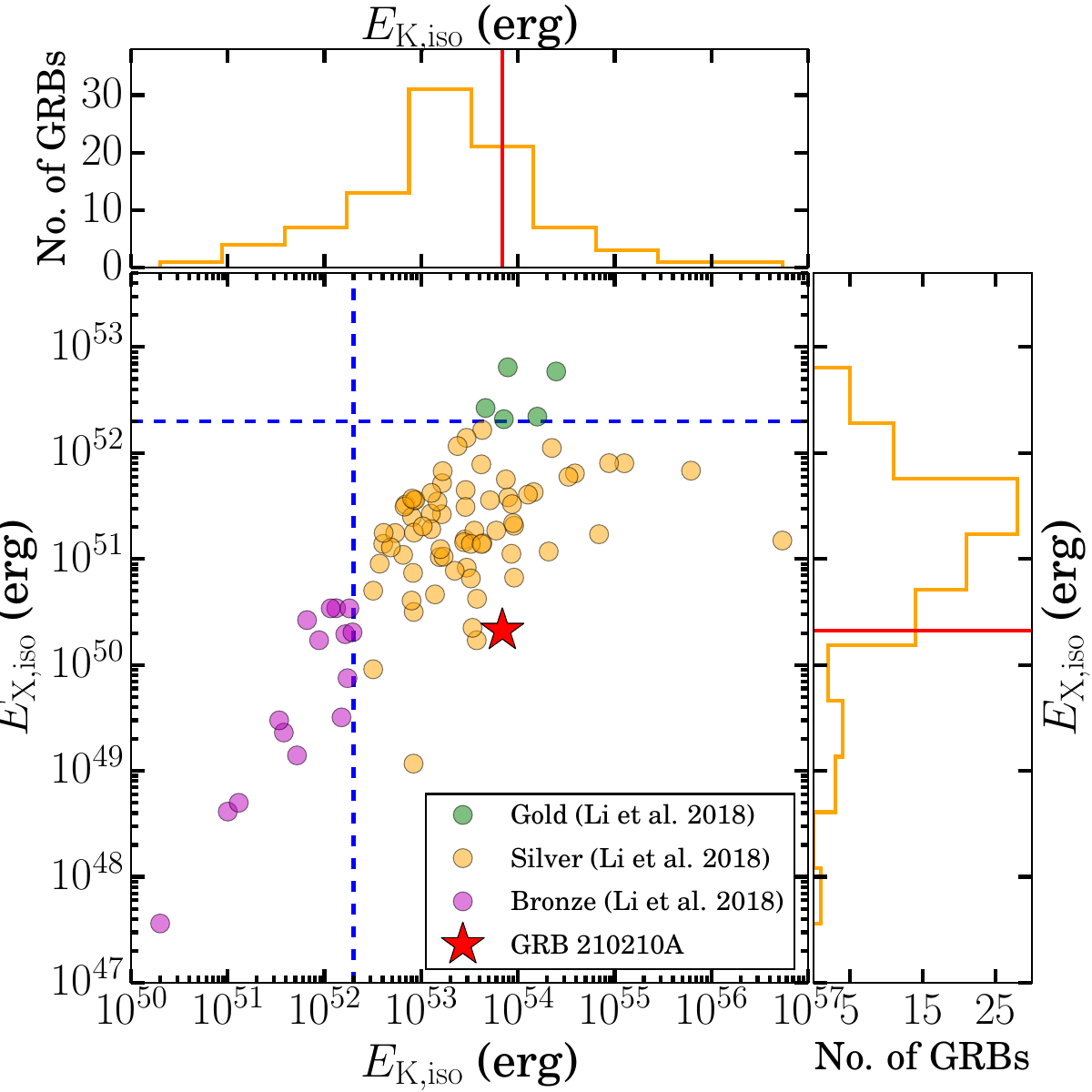}
\caption{The distribution of E$_{X, iso}$ and E$_{K, iso}$ for gold, silver, and bronze samples of \citet{2018ApJS..236...26L}, see also \citet{2024arXiv240601220R}. Dashed blue lines represent E$_{X, iso}$ and E$_{K, iso}$ values equal to 2 $\times$ 10$^{52}$ erg. GRB 210210A is shown with a red star. The histograms of observed E$_{X, iso}$ and E$_{K, iso}$ are shown in the right and upper panels, respectively. The position of GRB 210210A is shown with vertical red solid lines in these histograms.}
\label{fig:Ekiso}
\end{figure}

Furthermore, it might be the case that the observed early X-ray emission (within the 0.3-10 keV band) is not purely synchrotron emission coming from the external forward shock but also has a contribution from the central engine (plateau). In such a scenario, to maintain the level of the observed flux at a constant level, energy must be supplied to the fireball by the inner engine. We applied equation 7 from \citet{2018ApJS..236...26L} to calculate the X-ray energy during the early phase, finding it to be 2.12 $\times$ 10$^{50}$\,erg, which is within the total energy budget of a magnetar. Additionally, we used equation 11 from the same source to determine the kinetic energy (E$_{K, iso}$) of the fireball, which resulted in a value of 6.90 $\times$ 10$^{53}$\,erg. The measured E$_{X, iso}$ and E$_{K, iso}$ for \thisgrb are consistent with the silver sample from \citet{2018ApJS..236...26L}, supporting the magnetar central engine hypothesis for this burst (see Fig. \ref{fig:Ekiso}).

\subsubsection{Compact Object Mergers} 
Though less common, low-luminosity GRBs can also stem from the merger of compact binaries, such as the system of two neutron stars or a neutron star plus a black hole. These mergers may produce a weak jet, resulting in a fainter burst of gamma rays. The direct evidence of such progenitors is the detection of kilonova emission (optical/infrared transient due to the radioactive decay of heavy elements synthesized in the merger ejecta) and gravitational waves (potential detection of gravitational waves coincident with the gamma-ray signal if the event is relatively close). Given that such observations are challenging at the measured redshift of \thisgrb with current instruments, we aim to determine the origin (collapsar or merger) of \thisgrb following the indirect methodology of \citet{2011ApJ...739L..55B} and estimate its non-collapsar probability. The duration of the prompt emission of GRBs, represented by the \tninty value, must be at least as long as the engine remains active after the jet breakout. Typically, these two durations are considered equal in GRB models, expressed as \tninty = T$_{\rm Eng}$ - t$_{\rm b}$, where T$_{\rm Eng}$ is the duration of the engine's activity, and t$_{\rm b}$ is the time required for the jet to emerge from the surrounding envelope of the progenitor star. It is highly unlikely that the engine will stop operating exactly after the jet breakout. This condition is a fundamental aspect of the Collapsar model, implying that if \thisgrb originated from Collapsars, it must meet this criterion. We calculated t$_{\rm b}$ ($\sim$ 1.8 s) and the ratio T$_{\rm 90}$/t$_{\rm b}$ ($\sim$ 3.6) to identify the potential progenitor of \thisgrb, supporting the collapsar model. Furthermore, we estimated the non-collapsar probability of the burst using equations 2 and 3 from \citet{2013ApJ...764..179B}. The non-collapsar probability is 2.61 $\times$ 10$^{-3}$. Our analysis suggests that \thisgrb is more likely to be explained by instead of a merger of compact merger scenario.

\section{Summary and Conclusions}
\label{summary}

In this paper, we have analyzed the prompt and afterglow features of the intermediate luminosity GRB 210210A. This GRB is identified as one of the softest long-duration bursts detected by the \swift satellite. The prompt emission's time-integrated spectrum is effectively described by a power law with an exponential cutoff. The rest-frame spectral peak energy and the isotropic energy (E$_{\rm \gamma, iso}$) of GRB 210210A marginally satisfy by 2$\sigma$ of the Amati correlation, which is a typical feature of low to intermediate luminosity GRBs. This observed feature is common in low-luminosity GRBs. The observed characteristics of the prompt emission of \thisgrb indicate that it is a soft GRB, with \Ep $\sim$ 20 keV. Thus, GRB 210210A is an X-ray-rich burst belonging to the class of X-ray flashes \citep{2015PhR...561....1K}. The afterglow observations of this burst show chromatic behaviors. An early bump is observed in the X-ray light curve, which is a rare feature for GRBs. The optical light curve appears to follow a power law decay. However, due to insufficient early optical observations, the possibility of an early optical bump cannot be ruled out completely. We calculated parameters such as peak time, rise time, decay time, and bulk Lorentz factor ($\Gamma_{0}$ $\sim$ 156) for the early bump observed in the X-ray light curve. These calculated parameters perfectly satisfy the correlation found by \citet{2010ApJ...725.2209L} from the study of 17 optical and 12 X-ray afterglows with onset features. LLGRBs/intermediate luminosity GRBs may have jets with lower bulk Lorentz factors, resulting in less relativistic beaming and hence lower observed luminosity, which might be the case for \thisgrb. Both X-ray and optical afterglows exhibit a chromatic break (although these breaks are not well constrained due to fewer data points during these epochs) in the late afterglow phase, suggesting complex dynamics in the jet structure and surrounding medium. The parameters obtained by fitting the light curve reveal that the observed break is consistent with the jet break. We calculated the jet opening angle, breakout time, beaming corrected gamma-ray energy, kinetic energy, and X-ray energy during the early emission phase and suggested that the overall properties of GRB 210210A favor a collapsar scenario with a possible magnetar central engine. The analysis of GRB 210210A provides significant insights into the characteristics and behavior of intermediate luminosity GRBs. Further prompt observations of more LLGRBs/intermediate luminosity GRBs with soft X-ray instruments/missions (such as the {\it Einstein} Probe) and detailed theoretical studies are required to unravel the complexities of these less energetic events \citep{2023arXiv231216265G}. \\

\textbf{Acknowledgments:} We thank the referee for carefully reading the manuscript and the positive report. RG and MM were sponsored by the National Aeronautics and Space Administration (NASA) through contracts with ORAU. The views and conclusions contained in this document are those of the authors and should not be interpreted as representing the official policies, either expressed or implied, of the National Aeronautics and Space Administration (NASA) or the U.S. Government. The U.S. Government is authorized to reproduce and distribute reprints for Government purposes notwithstanding any copyright notation herein. RG and SBP acknowledge the financial support of ISRO under the AstroSat archival data utilization program (DS$\_$2B-13013(2)/1/2021-Sec.2). AJCT acknowledges support from the Spanish Ministry project PID2020-118491GB-I00 and Junta de Andalucia grant P20\_010168. AA acknowledges funds and assistance provided by the Council of Scientific \& Industrial Research (CSIR), India, under file no. 09/948(0003)/2020-EMR-I. AA also acknowledges the Yushan Fellow Program by the Ministry of Education, Taiwan for the financial support (MOE-111-YSFMS-0008-001-P1). This research has used data obtained through the HEASARC Online Service, provided by the NASA-GSFC, in support of NASA High Energy Astrophysics Programs. This work made use of data supplied by the UK Swift Science Data Centre at the University of Leicester.

\appendix

\onecolumn
\begin{longtable}{|c|c|c|c|c|c|c|} 
\caption{A log of photometric data for \thisgrb has been compiled, featuring observations from \swift UVOT, DFOT, and various GCN reports. It should be noted that the magnitudes listed have not been corrected for galactic or host galaxy extinction.}
\label{tab:UVOT}\\ \hline
FILTER & T$_{\rm start}$-\tzero (s) & T$_{\rm stop}$-\tzero (s) & MAG & MAG Error & Telescope & Reference\\ \hline
\endfirsthead
\hline 
\multicolumn{7}{|c|}{{Continued on next page}} \\ \hline
\endfoot
\hline
FILTER & T$_{\rm start}$-\tzero (s) & T$_{\rm stop}$-\tzero (s) & MAG & MAG Error & Telescope & Reference\\ \hline
 
\endhead
\hline
\endlastfoot

white & 85.1 & 234.9 & 16.05 & 0.03 & \swift UVOT & Present work \\ 
white & 578 & 597.8 & 16.34 & 0.06& \swift UVOT & Present work\\ 
white & 753.2 & 773 & 16.53 & 0.06& \swift UVOT & Present work\\ 
white & 877.8 & 1027.5 & 16.58 & 0.03& \swift UVOT & Present work\\ 
white & 1181 & 1373.8 & 16.68 & 0.05& \swift UVOT & Present work\\ 
white & 34748.6 & 35571.3 & 19.93 & 0.09& \swift UVOT & Present work\\ 
white & 58479.6 & 59166.3 & 20.36 & 0.21& \swift UVOT & Present work\\ 
white & 98188.5 & 98482.3 & $>$20.93 & - & \swift UVOT & Present work\\ 
white & 224611.5 & 224950.3 & $>$20.91 & - & \swift UVOT & Present work\\ 
white & 235162.2 & 367653.1 & $>$21.93 & - & \swift UVOT & Present work\\ 
white & 373259.7 & 373553.5 & $>$20.95 & - & \swift UVOT & Present work\\ 
white & 378869.7 & 385337.9 & $>$20.79 & - & \swift UVOT & Present work\\ \hline
b & 552.9 & 572.7 & 16.95 & 0.12& \swift UVOT & Present work\\ 
b & 728.8 & 748.5 & 16.83 & 0.11& \swift UVOT & Present work\\ 
b & 1156.5 & 1349.5 & 17.38 & 0.13& \swift UVOT & Present work\\ 
b & 33835.4 & 34743.5 & 19.88 & 0.15& \swift UVOT & Present work\\ 
b & 57567.7 & 98183.8 & $>$21.05 & -& \swift UVOT & Present work \\ 
b & 224268.1 & 367264.5 & $>$21.38 & -& \swift UVOT & Present work \\ 
b & 372961.4 & 373255.2 & $>$20.27 & -& \swift UVOT & Present work \\ 
b & 378676.6 & 385295.1 & $>$20.14 & - & \swift UVOT & Present work\\ \hline
u & 297.6 & 547.4 & 15.87 & 0.04& \swift UVOT & Present work\\ 
u & 703.1 & 722.9 & 15.97 & 0.1& \swift UVOT & Present work\\ 
u & 1131.8 & 1324.9 & 16.42 & 0.1& \swift UVOT & Present work\\ 
u & 17556.6 & 18366.4 & 18.95 & 0.1& \swift UVOT & Present work\\ 
u & 29044.6 & 29874.4 & 19.38 & 0.14& \swift UVOT & Present work\\ 
u & 97591.4 & 97885.2 & $>$19.97 & -& \swift UVOT & Present work \\ 
u & 223923.2 & 224262 & $>$19.93 & - & \swift UVOT & Present work\\ 
u & 234745.1 & 366875.5 & $>$21.01 & -& \swift UVOT & Present work \\ 
u & 372662.5 & 372956.3 & $>$20.01 & -& \swift UVOT & Present work \\ 
u & 378483.1 & 385251.9 & $>$19.84 & -& \swift UVOT & Present work \\ \hline
v & 628.6 & 648.4 & 16.88 & 0.23& \swift UVOT & Present work\\ 
v & 803 & 822.8 & 16.87 & 0.23& \swift UVOT & Present work\\ 
v & 1058.2 & 1250.5 & 16.85 & 0.17& \swift UVOT & Present work\\ 
v & 22352.7 & 23259.5 & 19.28 & 0.21& \swift UVOT & Present work\\ 
v & 52813.1 & 98766.5 & $>$19.57 & - & \swift UVOT & Present work\\ 
v & 224956.3 & 368106.6 & $>$20.33 & - & \swift UVOT & Present work\\ 
v & 373559.8 & 373866.5 & $>$19.3 & - & \swift UVOT & Present work\\ 
v & 379064.4 & 385386.5 & $>$19.17 & - & \swift UVOT & Present work\\ \hline
uvw1 & 678.8 & 698.6 & 16.13 & 0.16& \swift UVOT & Present work\\ 
uvw1 & 852 & 871.7 & 16.11 & 0.16& \swift UVOT & Present work\\ 
uvw1 & 1107.5 & 1300.4 & 16.55 & 0.14& \swift UVOT & Present work\\ 
uvw1 & 16650.3 & 17550 & 18.55 & 0.09& \swift UVOT & Present work\\ 
uvw1 & 28138.3 & 29038.1 & 19.13 & 0.13& \swift UVOT & Present work\\ 
uvw1 & 62967.5 & 63263.3 & $>$19.81 & -& \swift UVOT & Present work \\ 
uvw1 & 86691.3 & 93066.5 & $>$20.4 & -& \swift UVOT & Present work \\ 
uvw1 & 213506.8 & 213546.6 & $>$18.23 & -& \swift UVOT & Present work \\ 
uvw1 & 217934.1 & 218166.5 & $>$19.6 & - & \swift UVOT & Present work\\ \hline
uvw2 & 603.4 & 623.2 & 16.55 & 0.2& \swift UVOT & Present work\\ 
uvw2 & 778.6 & 798.3 & 16.3 & 0.18& \swift UVOT & Present work\\ 
uvw2 & 1033.5 & 1226.1 & 16.53 & 0.14& \swift UVOT & Present work\\ 
uvw2 & 1379.5 & 1386.3 & $>$16.63 & -& \swift UVOT & Present work \\ 
uvw2 & 51906.9 & 52806.6 & $>$20.6 & -& \swift UVOT & Present work \\ 
uvw2 & 85572.6 & 92105.9 & $>$20.8 & - & \swift UVOT & Present work\\ 
uvw2 & 161135.1 & 161288.9 & $>$19.43 & -& \swift UVOT & Present work \\ 
uvw2 & 213350.1 & 213423.9 & $>$18.84 & -& \swift UVOT & Present work \\ 
uvw2 & 217416.6 & 217670.3 & $>$19.83 & -& \swift UVOT & Present work \\ \hline
uvm2 & 652.9 & 672.7 & 16.3 & 0.22& \swift UVOT & Present work\\ 
uvm2 & 827.3 & 847.1 & 16.06 & 0.2& \swift UVOT & Present work\\ 
uvm2 & 1082.6 & 1275 & 16.48 & 0.17& \swift UVOT & Present work\\ 
uvm2 & 23264.7 & 23526.3 & 19.22 & 0.29& \swift UVOT & Present work\\ 
uvm2 & 86131.8 & 92645 & $>$20.61 & -& \swift UVOT & Present work \\ 
uvm2 & 161293.6 & 161346.6 & $>$18.22 & -& \swift UVOT & Present work \\ 
uvm2 & 213428.5 & 213502.3 & $>$18.56 & - & \swift UVOT & Present work\\
uvm2 & 217675.4 & 217929.2 & $>$19.58 & - & \swift UVOT & Present work\\ \hline
\multicolumn{7}{|c|}{}\\ \hline
FILTER & T-\tzero (s) & Exp time (s) & MAG & MAG Error & Telescope & Reference\\ \hline
R & 335989.00 & 15$\times$180 & $>$21.8 & - & DFOT & Present work\\

\hline
\multicolumn{7}{|c|}{Data taken from GCN circular} \\ \hline
FILTER & \multicolumn{2}{c|}{T-\tzero (days)} & MAG & MAG Error & \multicolumn{2}{c|}{Reference}\\ \hline
R & \multicolumn{2}{c|}{0.0674} & 17.60 & 0.20 &\multicolumn{2}{c|}{\citep{2021GCN.29469....1J}}\\
R & \multicolumn{2}{c|}{0.1763} & 18.84 & 0.03 &\multicolumn{2}{c|}{\citep{2021GCN.29450....1D}}\\
R & \multicolumn{2}{c|}{0.3967} & 19.62 & 0.05 &\multicolumn{2}{c|}{\citep{2021GCN.29459....1S}}\\
R & \multicolumn{2}{c|}{0.8800} & 21.35 & 0.07 &\multicolumn{2}{c|}{\citep{2021GCN.29488....1D}}\\
R & \multicolumn{2}{c|}{1.1674} & 21.89 & 0.21 &\multicolumn{2}{c|}{\citep{2021GCN.29476....1K}}\\
R & \multicolumn{2}{c|}{3.1294} & 23.05 & 0.14 &\multicolumn{2}{c|}{\citep{2021GCN.29502....1K}}\\

\end{longtable}

\begin{table*}[!ht]
\centering
\caption{Parameters obtained from the fitting \swift-UVOT light curve with power law and broken power law model.}
\label{tab:model}
\begin{tabular}{|c|c|c|c|c|} \hline
Filter & $\alpha_{1}$ or  $\alpha$ & t$_{b}$ & $\alpha_{2}$ & $\chi^{2}_{\nu}$\\ \hline
\multicolumn{5}{|c|}{Power law}\\ \hline
b  &  \multicolumn{3}{c|}{-0.68$_{-0.04}^{+0.04}$} & 2.01\\ 
u  &  \multicolumn{3}{c|}{-0.76$_{-0.02}^{+0.02}$}& 4.75\\ 
v  &  \multicolumn{3}{c|}{-0.70$_{-0.07}^{+0.07}$} & 1.58\\ 
uvw1  &  \multicolumn{3}{c|}{-0.74$_{-0.03}^{+0.03}$} & 0.72\\ 
uvm2  &  \multicolumn{3}{c|}{-0.86$_{-0.01}^{+0.01}$} & 1.57\\ 
uvw2  &  \multicolumn{3}{c|}{-0.87$_{-0.09}^{+0.08}$} & 5.22\\ \hline
\multicolumn{5}{|c|}{Broken Power law}\\ \hline
White  &  0.28$_{-0.02}^{+0.02}$ & 1545.84$_{-303.11}^{+303.46}$ & 0.94$_{-0.05}^{+0.05}$ & 0.87\\\hline
\end{tabular}
\end{table*}

\end{document}